\documentclass[usegraphicx,usenatbib,useapjfonts,apj]{emulateapj}

\def\plotancho#1{\includegraphics[width=18cm]{#1}}

\slugcomment{To appear in the Astronomical Journal}

\shorttitle{New morphological classification of the CIG catalog based upon the SDSS}

\shortauthors{Hern\'andez-Toledo et al.}

\begin{document}

\title{A morphology revaluation of galaxies in common from the Catalog of Isolated 
Galaxies and the Sloan Digital Sky Survey (DR6)}

\author{H. M. Hern\'andez-Toledo\altaffilmark{1}, J. A. V\'azquez-Mata\altaffilmark{2}, 
L. A. Mart\'{\i}nez-V\'azquez, V. Avila-Reese, \\
H. M\'endez-Hern\'andez, 
S. Ortega-Esbr\'{\i}, J. P. Miranda-N\'u\~nez}

\affil{Instituto de Astronom\'{\i}a, Universidad Nacional Aut\'onoma de M\'exico, A.P. 70-264, 04510 
M\'exico D. F., M\'exico.}

\altaffiltext{1}{E-mail: hector@astroscu.unam.mx}
\altaffiltext{2}{E-mail: jvazquez@astroscu.unam.mx}

\begin{abstract}      

We present a revaluation of the optical morphology for 549 galaxies from the Catalog of 
Isolated Galaxies in the Northern Hemisphere (CIG; Karachentseva 1973) that are available in 
the Sloan Digital Sky Survey (SDSS; DR6). Both the high resolution and high dynamic range of the 
SDSS images and our semi--automatic image processing scheme, allow for a major quality and uniform 
morphological analysis. The processing scheme includes (1) sky--subtracted, 
cleaned, and logarithmic scaled $g-$band images; (2) filtered--enhanced versions of the images 
in (1); and (3) the corresponding RGB images available in the SDSS database. Special attention 
was paid for distinguishing between E, S0, and Sa candidates through an additional analysis of 
(4) the surface brightness, $\epsilon$, $PA$ and $A_{4}/B_{4}$ Fourier series expansion profiles.  
An atlas of mosaics containing [(1), (2) and (3)] images for Sab--Sm/Irr types and 
[(1), (2), (3), (4)] images for E/S0/Sa types was produced and is available on the web.
The median type in the sample corresponds to Sbc, with 65\% of the sample being of this
type or later. A scarce population of E (3.5\%) and S0 (5\%) galaxies amounting 
to 8.5\%, and a non-negligible 14\% fraction of early--type SaSab spirals are identified. 
We compare our results against a previous reclassification 
of the CIG catalog based on the digitized POSS II images (Sulentic et al. 2006).  We calculate
also the $gri$ absolute magnitudes corrected by Galactic and internal extinctions and present the 
$g-i$ color distribution and the color--magnitude diagram. 
Among the spirals, we find tentative fractions of strong and suspected bars of 65.8\% and 
of 33.3\% of rings. A detailed image analysis of the E galaxies (18) allows us to find a richness 
of distinct substructure in their isophotal shape and also of morphological distortions.
At least 78\% of the E galaxies show some kind of morphological distortion (shells, dust
lanes,  diffuse halos, etc.), suggesting that these galaxies suffered late dry mergers.
The isophotes of 42\% (37\%) of the E galaxies are boxy (disky). Among 4 blue E's,
3 are disky. 
Finally, we calculate for all the sample the $CAS$ (concentration, asymmetry and 
clumpiness) structural parameters in the $ugriz$ bands.  We analyze the loci of these galaxies 
in different projections of the $CAS$ volume diagram and discuss some trends of 
the $CAS$ parameters with the color band, as well as with the morphological type and the
galaxy color.
\end{abstract}


\section{Introduction} \label{S1}

A galaxy is considered as isolated if it has not suffered any interaction with 
another normal galaxy or with a group/cluster environment over a Hubble time
or at least since approximately one half of its mass was assembled. In this
sense, isolation offers a unique opportunity to identify  
the intrinsic physical and evolutive processes of galaxies, and then to confront
the results with theoretical predictions of galaxy evolution (e.g., Avila-Reese
et al. 2008). Isolated galaxies
are also important as comparison objects in studies of the morphological mix 
as a function of environment, and the environmental effects on galaxy properties. 
However, the definition of isolated galaxy from an observational
point of view is not an easy task. Karachentseva (1973) introduced an operational
criterion for this definition and constructed the Catalog of Isolated Galaxies
in the Northern Hemisphere (hereafter CIG), one of the most extended and used
samples for these objects. The sample contains 1050 galaxies of all morphological
types and $m_B<15.7$ mag, identified in the photographic Palomar Observatory 
Sky Survey (POSS) plates. A careful revision of 
the morphology and general properties of these galaxies in the light of recent, 
much more detailed observations is highly desirable.

A first step in this direction has been given by Sulentic et al. (2006; hereafter
Sul06), who
revisited the morphology of the CIG galaxies by using the digitalized POSS II data,
and found some fraction of galaxies having a morphological type different to those
reported in the original CIG.   In this paper, we are aimed to analyze the CIG galaxies
in common with the huge galaxy sample of the Sloan Digital Sky Survey (SDSS),
a sample that we will call CIG$\cap$SDSS.
The advantages of the SDSS CCD images over previous image surveys in similar regions 
of the sky are notorious. Among these we mention: (1) the uniformity in both the 
instrumental set and photometric conditions, (2) the homogeneity in multicolor 
$ugriz$ imaging and photometry, and (3) the high resolution  (0.4 $\arcsec$/pix) of 
the images.  These advantages are crucial for studying galaxy morphology and structural
properties.

Because of the SDSS large numbers, uniformity, and completeness down to a certain magnitude, 
it is an ideal database for the search of fundamental relations between physical 
properties of galaxies and their dependencies on local and larger--scale environments. 
For example, recently 
Park et al. (2007,2008) used the SDSS data for studying the environmental dependence 
of the observed morphology --and other properties-- at different density smoothing scales. 
Based on their results, they concluded that galaxy morphology depends primarily
on the small scale environment, which may reflect the influence of later evolutionary 
effects such as galaxy--galaxy interactions, rather than on large--scale environment 
in which the galaxy initially formed.  A relevant goal of the present paper is the identification of the
isolated galaxies that may show evidence of morphological disturbances. It is possible that the isolated galaxies 
showing signs of distortions are the recent merger products, at a given large-scale background density, as 
predicted by Park, Gott \& Choi (2008).

An important question concerning morphological classification is the correct definition of
what is an elliptical (E), lenticular (S0) or early--type spiral (Sa) galaxy. 
These types are the most affected in their definition by saturation, poor resolution, 
etc. Saucedo--Morales \& Bieging (2001) estimated that one--half of E galaxies could be 
misclassified in the POSS plates--based CIG catalog. 
The existence of true isolated early--type galaxies is in general an important  
problem in galaxy formation. Our finding of detailed structure in the E and S0 
isolated galaxies is expected to provide new tests for the various hypotheses of their 
formation. Have the isolated E galaxies systematic differences from those ones 
found in groups and clusters? Have they assembled in the same way in both cases?   
Similar questions could be asked about the isolated S0 galaxies, whose 
detailed formation histories are poorly understood.   
A few decades back, Gott \& Thuan (1976) argued that E galaxies were 
produced by larger initial--density fluctuations, which would have higher density 
at turnaround and where star formation (SF) would be completed before collapse. 
Such larger initial--density fluctuations would be more likely in a region that 
would later turn into a high density environment. In the modern context of the
$\Lambda$ Cold Dark Matter ($\Lambda$CDM) hierarchical model, the formation of E/S0 galaxies
is also favored in the dense environments. The models predict a small probability
of finding isolated E/S0 galaxies and they should have younger stars on average
than their cluster counterparts, as well as assembled their present--day 
masses by late major mergers (Kauffmann 1996; de Lucia et al. 2006). 

A complementary analysis of galaxy morphology and gross photometric
properties has been introduced some years ago, namely the $CAS$ (concentration, asymmetry 
and clumpiness) system. This system has been proposed to distinguish galaxies at 
different stages of evolution (Conselice 2003, and references therein) and it
provides the possibility to carry out a physical and quantitative classification 
based on measurable structural parameters. We present here the measured $CAS$ parameters
in the $ugriz$ bands for the 549 CIG$\cap$SDSS isolated galaxies. 
 
The outline of the paper is as follows. Section 2 summarizes the relevant past work on 
CIG morphologies. In \S 3 the image processing scheme and considerations used to reevaluate 
morphology for the CIG$\cap$SDSS galaxies are presented. In \S 4 a description of our 
Atlas of mosaic images for the 549 CIG$\cap$SDSS galaxies is presented.  The results of 
our classification and morphological analysis, and the calculation of the $CAS$  parameters 
in the $ugriz$ bands are presented. Section 5 is a discussion containing  a comparison of our morphological 
classification to previous works (\S 5.1), a discussion on the bar and ring fractions in spiral galaxies (\S 5.2), the morphological
distortions and formation of isolated E galaxies (\S 5.3), and the $CAS$ parameters
in different bands and their correlations with other galaxy properties (\S 5.4). 
Finally, in \S 6 a summary and the main conclusions of this paper are presented.

\section{CIG Morphology in the Literature} \label{DataSample}

The morphological classification as well as the environment of each CIG galaxy was originally inspected using the Palomar 
Observatory Sky Survey (POSS I) images which were based on  Kodak blue 103aO and red 103aE emulsions. According to Sulentic (1989), 
the representative numbers in the CIG were 168 E/S0 and 
883 Sa-Sm/Irr.  The morphological refinement of the CIG  was gradually 
carried out as new-quality data in the optical and other wavelengths were accumulated since the POSS I era. As the result,  for example 
about one-half of E's in the original CIG were found to be misclassified (Saucedo-Morales \& Bieging 2001).
A detailed review on previous works about the morphological refinement for the CIG 
galaxies is out of the scope of this paper; for a recent review on this see Sulentic et 
al. (2006; hereafter Sul06). The role of extragalactic databases in collecting and 
homogenizing data of diverse nature (from photographic, photoelectric and CCD images) 
has been fundamental. We mention the special role played by NED and HyperLeda databases,  which 
will be used extensively through this work.

The morphological content of the whole CIG sample has been reevaluated by Sul06, who 
used the digitalized POSS II data. POSS II is based 
on the IIIaJ and  IIIaF emulsions providing higher contrast and resolution than the previous POSS I. Their main results concerning 
the morphological content are: (1) 82\% of the CIG galaxies are spirals (Sa-Sd) with the bulk being luminous systems with small 
bulges (63\% are between Sb and Sc types), and  (2) 14\% are early-type E/S0 galaxies (the remaining 4\% are suspected to be irregular 
galaxies. Sul06 also reported a 
considerable number of  galaxies in the CIG catalog (n = 193) flagged for the presence of nearby companions or signs of distortion 
likely due to interactions.  Those results represent a more uniform and homogeneous reclassification for the whole CIG catalog and 
provide an excellent frame to test the results of our new morphological revaluation for 549 CIG galaxies with images available in 
the SDSS (DR6) database. 

Recent examples of detailed morphological reclassifications for CIG galaxies based on optical CCD images, from San Pedro M\'artir 
National Observatory, and extended to the Near-Infrared domain can be found in  Hern\'andez-Toledo et al. (2007; 2008). H$\alpha$ 
morphological studies in another subsample of CIG isolated spiral galaxies have been reported by Verley et al. (2007).

\section{The CIG$\cap$SDSS Data Sample and Image Processing} \label{DataSample}

\subsection{Isolated Galaxies from the SDSS DR6. The Data}

The SDSS is a digital photometric and spectroscopic survey obtained with a mosaic camera that images the sky by scanning 
along great circles at the sidereal rate (Gunn et al. 1998). The imaging data are produced simultaneously in the 
photometric $u, g, r, i$, and $z$ bands, with effective wavelength bands of 3551, 4770, 6231, 7625, and 9134 $\AA$, 
respectively (Fukugita et al. 1996) under photometric conditions (Hogg et al. 2001). The SDSS data are reduced 
by highly automated photometric and spectroscopic reduction pipelines (see Stoughton et al. 2002). Sources are 
identified, deblended, and photometrically measured (Lupton et al. 2002), and then the magnitudes are calibrated
to a standard star network approximately in the AB system (Smith et al. 2002). Very recently the SDSS team 
made public their DR6 to the astronomical community. SDSS DR6 covers 9583 $deg^{2}$ of five-band imaging data of the sky
and includes spectra, with derived spectroscopic parameters, for 68 770 stars, 790 860 galaxies, and 103 647 quasars. Based on 
NED and HyperLeda coordinates for the whole CIG sample, all the available SDSS $ugriz$ band images (scale 0.396 
$\arcsec$/pixel) were retrieved from the SDSS archive for a total of 549 entries from CIG (the
CIG$\cap$SDSS sample).  

 \begin{figure*}
\vspace{19.1cm}
\hspace{14.5cm}
\includegraphics{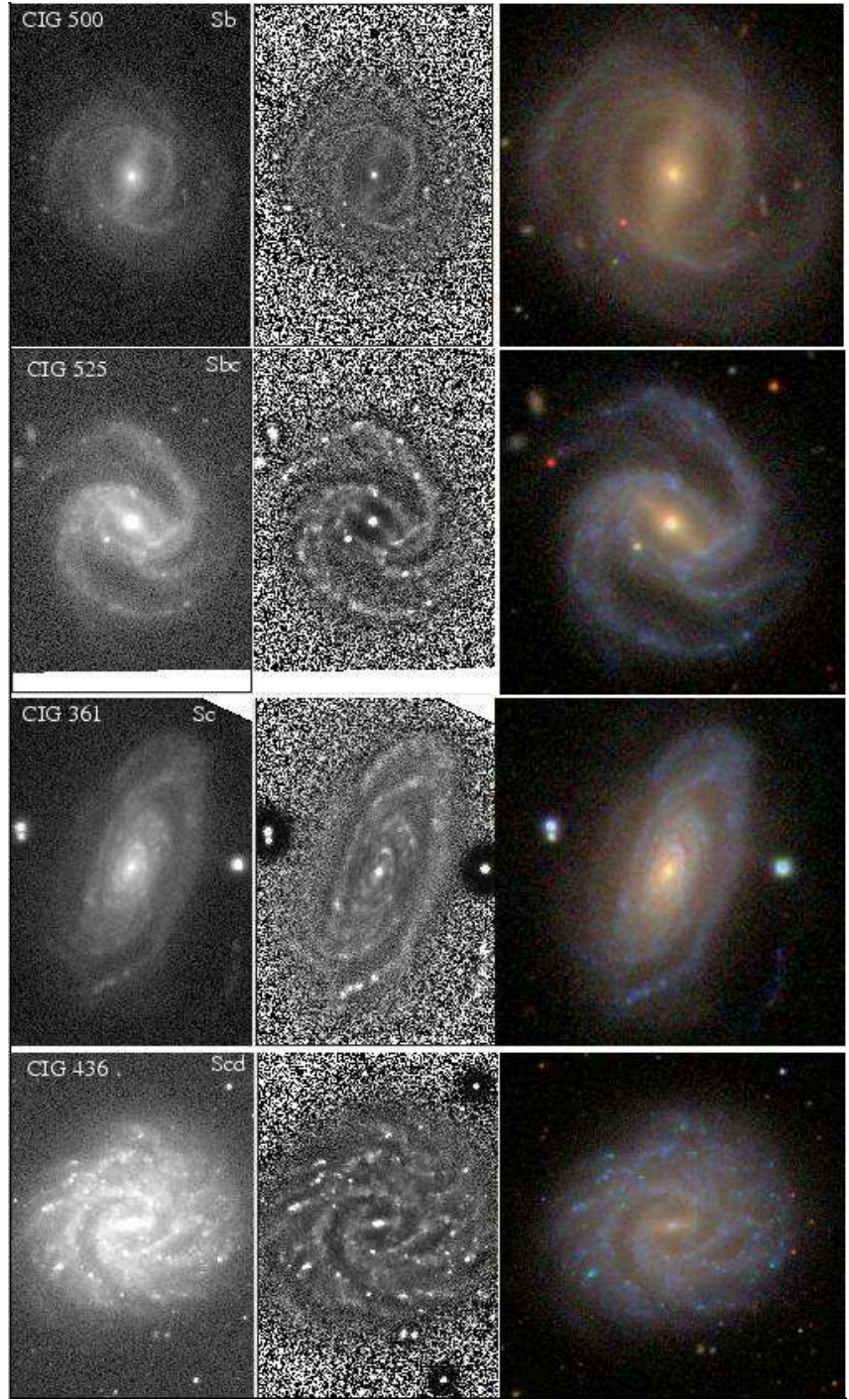}
\caption{CIG 500: a spiral galaxy with a prominent bulge classified as SBb(r). CIG 525: 
the bulge is less prominent and the arms become to show more definite structure. Classified as SBbc(r). CIG 361: the bulge is even less prominent and the 
arms show significant fragmentation. Classified as SABc.  CIG 436: a late-type spiral with clear fragmentation/resolved structures in the arms. Classified as SBcd(r).}
\label{mosaico1}
\end{figure*} 

To process the images from the SDSS (DR6) a procedure  that i) efficiently detects stars, ii) finds their positions, iii) estimates 
the nearby sky background and iv)  replace the stars in the images by a close representative background, was implemented.  
For this purpose several routines from the IDL Daophot package were used. To improve the performance of these routines, the 
signal to noise ratio in the images was enhanced by suppressing high spatial frequency image noise. This task was accomplished 
by using a small digital low-pass filter (Mighell 1999). After identifying the position of a target galaxy, a next stage starts by creating 
a mask of about two times the corresponding Petrosian radius above each galaxy. The covering area around a target galaxy was 
set to protect and preserve the local characteristics of each galaxy image while cleaning the rest of the image of field stars. At the 
same time, the image sections containing detected stars were substituted by local background values. Mean background is 
restimated and subtracted from the whole new image and optionally from a section of the image, depending on the position of the 
target galaxy within the original CCD frame. The procedure was completely programed within the IDL platform by using the 
routines provided by the IDL Astronomy User's Library which includes Daophot. Through this procedure we were able to process 
thousands of images and repeat the image procedure to find,  after trial/error tests, the optimal parameters related to the star subtraction 
process. 

In the next stage, all the images were automatically filtered through a mean Gaussian kernel and at the same time, the diffuse 
background was canceled out following filter-enhancing techniques (Sofue 1993). A three-component mosaic containing (1) a 
clean/sky subtracted and logarithmic scaled $g-$band image, (2) the corresponding filtered-enhanced version of (1) and (3) the 
corresponding RGB image, available in the SDSS database were used for each individual galaxy. As a first step of the morphological 
classification we sorted the sample into E/S0/Sa and Sab-Sm/Irr candidates. Next, different kernel values were used to filter and 
enhance different morphological details in early and late-type galaxies to further proceed with a second classification stage. The 
resultant images were used to create 1) three-component image mosaics for Sb-Irr galaxies and 2) three-component images plus 
the corresponding surface brightness and  geometric profiles after an isophotal analysis for E/S0/Sa galaxies.

 \begin{figure*}
\vspace{19cm}
\hspace{14.5cm}
\includegraphics{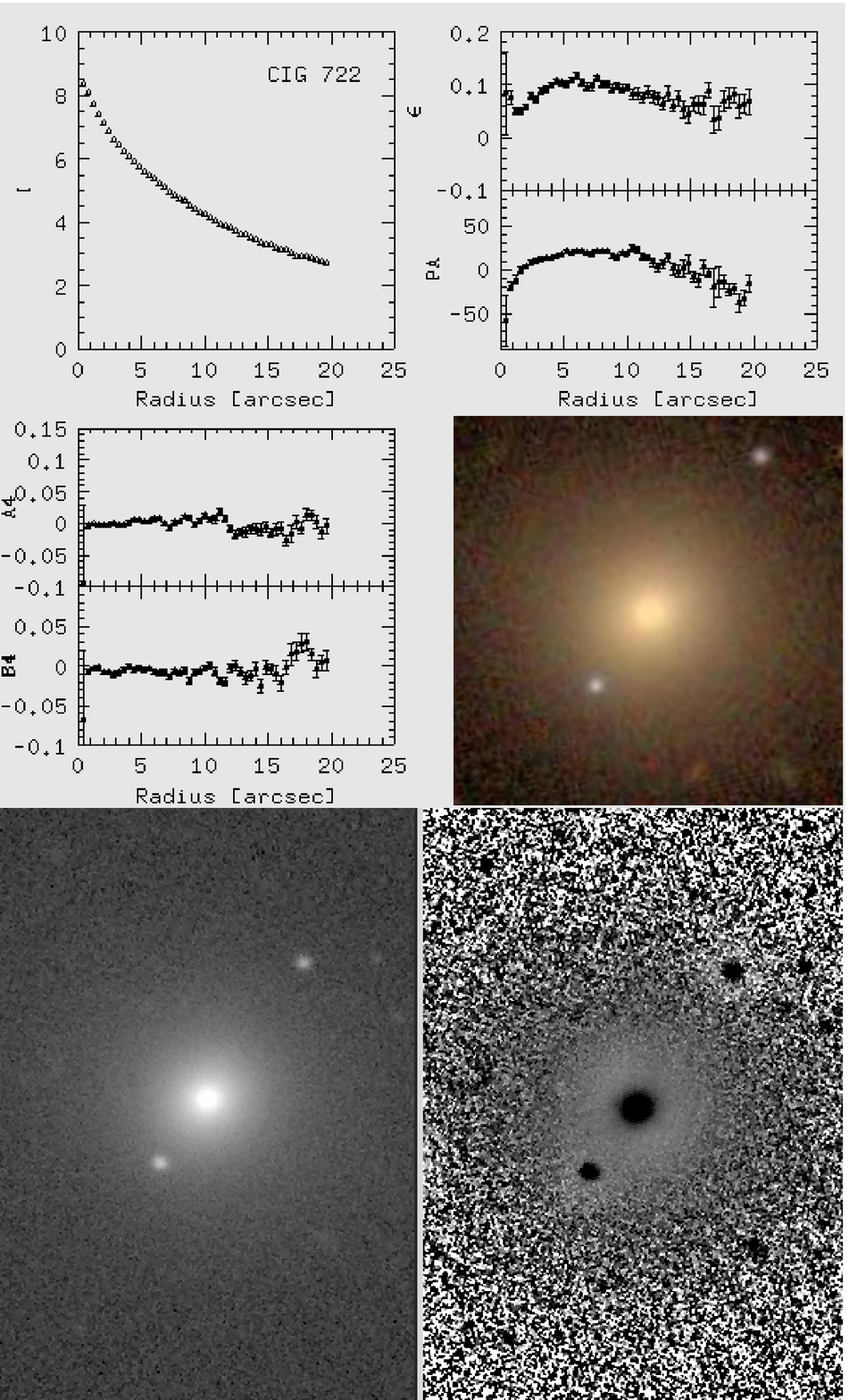}
\caption{CIG 722: a prototype of an elliptical galaxy. The upper panels show the Surface brightness, the ellipticity $\epsilon$, Position Angle $PA$ radial profiles and $A_{4}/B_{4}$ coefficients of the Fourier series expansions of deviations of a pure ellipse  in the $g$- band. The lower panels show 
the RGB color map form SDSS, the logarithmic-scaled $g$-band image and its filtered-enhanced version. }
\label{mosaico1}
\end{figure*}

\subsection{Morphological Considerations}

Galaxy classification is strongly dependent on the quality of the data employed (s/n ratio and resolution) and of course, on the 
wavelength. We state here our general criteria to carry out such a classification in the $g$ band. We expect to find more detailed 
and reliable structure because of the improved scale and dynamic range of the SDSS database. In order to discuss the optical 
morphology and its relationship to the global photometric properties, we present the 549 
CIG$\cap$SDSS mosaics, each one including 1) a gray scale 
$g$-band image displayed in logarithmic scale to look for internal/external details; 2) an $g$ band filtered-enhanced image to look 
for internal structure in the form of star forming regions, bars, rings and/or structure embedded into dusty regions; for E's this 
method is useful to reveal fine structure (e.g. dust, shells, broad tides, etc.) underneath the dominant light of the galaxy;  and 3) an 
RGB color image from the SDSS database to visualize the spatial distribution of the SF and other components like dust (blue colors 
for recent SF and red colors for older populations/dusty components) to complement the structural and morphological analysis.

In the case of E/S0/Sa candidates, besides including the processed images, we use the ISOPHOTE package in IRAF to fit ellipses 
to the galaxy distribution of light.  This package uses an iterative method that starts at a small radius and increases to large radii in a 
geometrical progression. The galaxy center, position angle and ellipticity were allowed to vary. These isophotal shape parameters 
help define the morphology of the galaxy. For example, the fourth-order cosine term of the Fourier series is a useful parameter to 
express the deviation from a perfect ellipse due to the presence of additional light. An excess of light along the major and/or minor 
axes (disky) is indicated by positive values, while negative values indicate excess light at $45^{\o}$  with respect to these axes (boxy).  
Each mosaic includes a surface brightness profile and the corresponding geometric profiles of ellipticity ($\epsilon$), Position Angle 
($PA$) and $A_{4}/B_{4}$ coefficients of the Fourier series expansions estimated from the $r$ band images.

\subsubsection{Sab--Sm/Irr Types}

The classification of the sample follows the basic Hubble sequence. For S types we considered the bulge to disk ratio as judged 
from the observed prominence of the bulge, tightness of the arms, and the degree of resolution of structure along the arms/outer disk. 
In the majority of the CIG spirals these features are well recognized, however in some cases the presence of structures like dust lanes, 
prominent knots and the apparent tightening of the arms in the central regions may confuse the identification of inner rings or bars. 
Outer rings/pseudo-rings (Buta 1995) were also identified when possible.  While the presence/absence of a bar was confirmed in 
some cases (B), in the suspected cases we adopted the (AB) nomenclature convention. In most of the cases when the inclination of 
the galaxy is greater than $80^{o}$ (see Table1) we tend to adopt the classification from the literature.

Figure 1 illustrates our image procedures and shows 4 S galaxies (CIG 500, CIG 525, CIG 361 and CIG 436) that where classified 
according to the stated criteria. Notice how the bulge is relatively decreasing in importance and the arms become significantly 
fragmented/resolved into clumps indicating, from top to bottom, Sb, Sbc, Sc, and Scd types. Some main structural features like bars and 
rings were identified and sometimes suspected. Each galaxy is identified by its CIG number and the corresponding morphological type.
 

The first row of Figure 1 shows CIG 500 displayed in logarithmic scale (left-panel). The prominence of the bulge/bar region and at the 
same time the smoothness of a not so tightly wounded set of arms are appreciated. The $g$-band filtered-enhanced image (middle 
panel) let us appreciate more detailed structure along the bar, ring, and arms. The filtered-enhancing techniques (Sofue 1993) allow the 
subtraction of the diffuse background in a convenient way to discuss different morphological details. The RGB composed image (right 
panel) from the SDSS database complement and reinforces our view of a prominent bulge/bar region and the smoothness of the arms. 
While HyperLeda classifies this galaxy as $Sab$ NED reports an $(R')$SAB$(r)$ab type with an outer pseudo-ring and an internal ring. 
The most recent revaluation by Sul06 classifies CIG 500 as SBb. We classify this galaxy as SBb$(r)$. 

The second row in Figure 1 shows CIG 525. The logarithmic image shows a slightly less prominent bulge but a significant bar and 
moderately opened arms. The $g$-band filtered-enhanced image shows more detailed structure along the arms and the inner ring. The 
RGB image complements our view emphasizing detailed structure along the arms. While HyperLeda, NED and Sul06 classify this galaxy 
as Sb, we classify CIG 525 as SBbc$(r)$.  

The third row of Figure 1 shows CIG 361. The logarithmic image shows a galaxy with a less prominent bulge but with a set
of quite opened arms showing  significant fragmentation. The filtered-enhancing image let us suspect about a central 
barred region and emphasizes the fragmentation along the arms. The RGB SDSS image clearly supports a detailed view of 
all these features. While HyperLeda and NED classify this galaxy as Sb, Sul06 classifies CIG 361 as Sab. In contrast, 
and from the observed structure, we classify CIG 361 as SABc. 
 
Finally the fourth row in Figure 1 shows CIG 436. The bulge appears not so significant and the arms are strongly fragmented/resolved 
into clumps. An inner ring and a bar region are readily appreciated. While HyperLeda classifies this galaxy as SBc, NED, Sul06 and 
we classify CIG 436 as SBcd. A more detailed classification from our images is SBcd(r).

\begin{figure*}
\vspace{19.1cm}
\hspace{14.5cm}
\includegraphics{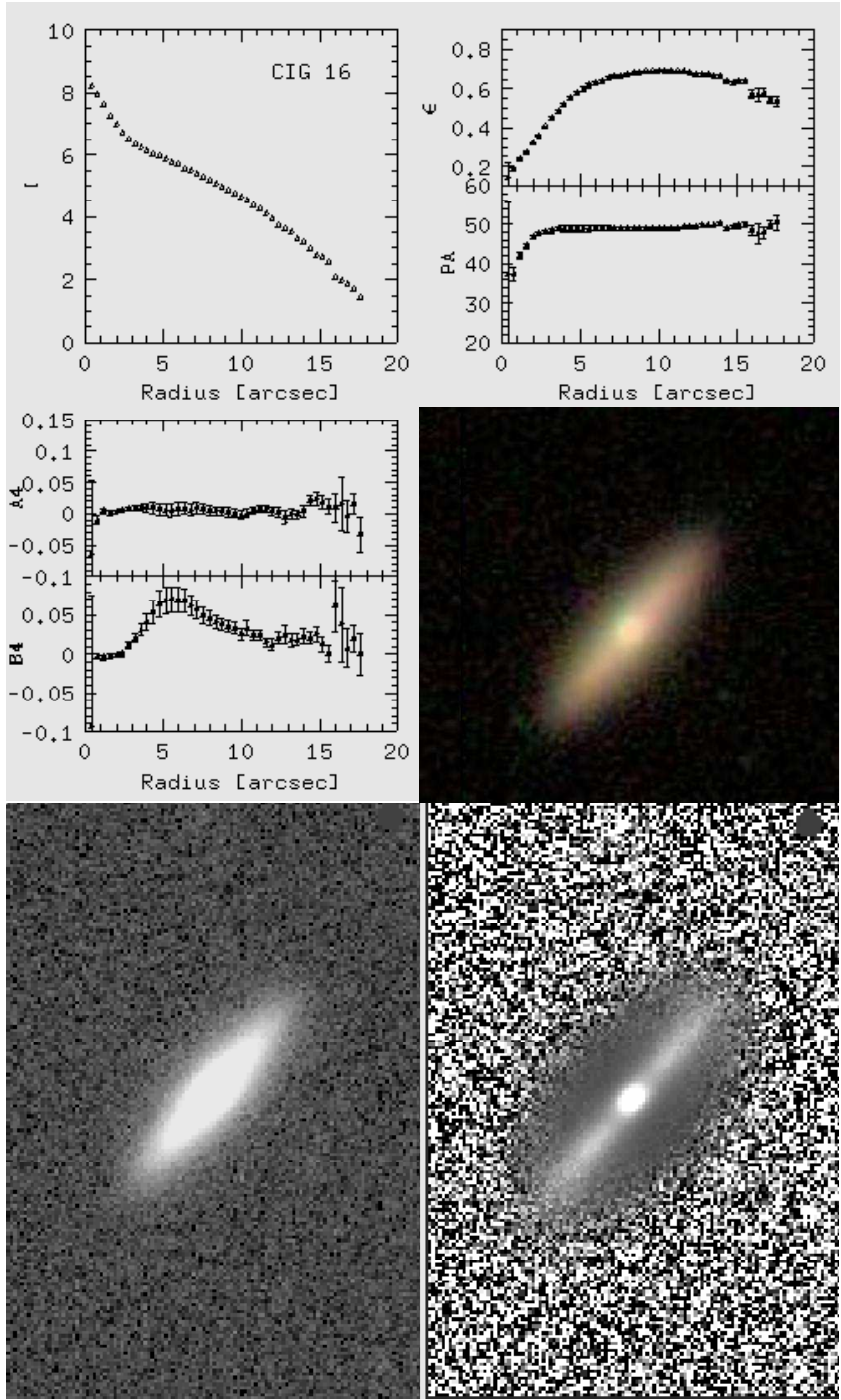}
\caption{CIG 16: a prototype of a lenticular galaxy. Same as in Figure 2.}
\label{mosaico1}
\end{figure*}

\subsubsection{E/S0 Types}

For early-type (E/S0/Sa) candidates, besides a careful inspection to the processed images, we carried out an evaluation of the 
geometrical parameters after an isophotal analysis. The ellipticity $\epsilon$, position angle $PA$, coefficients of the Fourier series 
expansions $A_{4}$, $B_{4}$ and the surface brightness radial profile are also presented for each galaxy.  Although the absolute value 
of the $A_{4}$ parameter depends on the inclination of the galaxy to the line of sight, its sign is useful to detect subtle disky features in 
the early-type candidates. The presence of disky isophotes is one criterion for assigning a galaxy to the lenticular class. However, disks 
or disky components have also been found among the E class (e.g., Bender 1988; Nieto et al. 1988). A galaxy was judged to be E 
if the $A_{4}$ parameter showed: 1) no significant  boxy ($A_{4} < 0$) or disky ($A_{4} > 0$) trend in the outer parts, or 2) a generally boxy 
($A_{4} < 0$) character in the outer parts and if in addition, the surface brightness profile showed 3) the absence of a linear component 
in the surface brightness profile. Central diskiness is considered not enough for an S0 classification. Additionally we paid attention to 
changes in the $\epsilon$ and $PA$ radial profiles.

For an E type we should not observe a significant change in both $\epsilon$ and $PA$, while in the case of an S0, more significant 
changes in these parameters could be showing evidence of a disks. The presence of structure in the outer regions of E's like 
rings/shells and diffuse haloes (DH)  could mimic an external disk as viewed from the point of view of the geometric profiles. However, if 
our additional image processing did not show evidence of definite features, we kept the galaxy type as E. The surface brightness 
radial profile is also an important tool to discriminate between E/S0 types. It is known that in the E's the surface brightness profile 
follows either a de Vaucouleurs or a more general S\'ersic law. When the profile presents a significant deflection in the curvature into 
a linear regime (not associated to other internal structures like inner disks, rings or bars), this could be evidencing the presence of an 
external disk. Figures 2 and 3 illustrate our imaging procedures for identifying E and S0 galaxies.  Each galaxy is identified by its CIG 
number. 


\begin{figure*}
\vspace{19cm}
\hspace{14.5cm}
\includegraphics{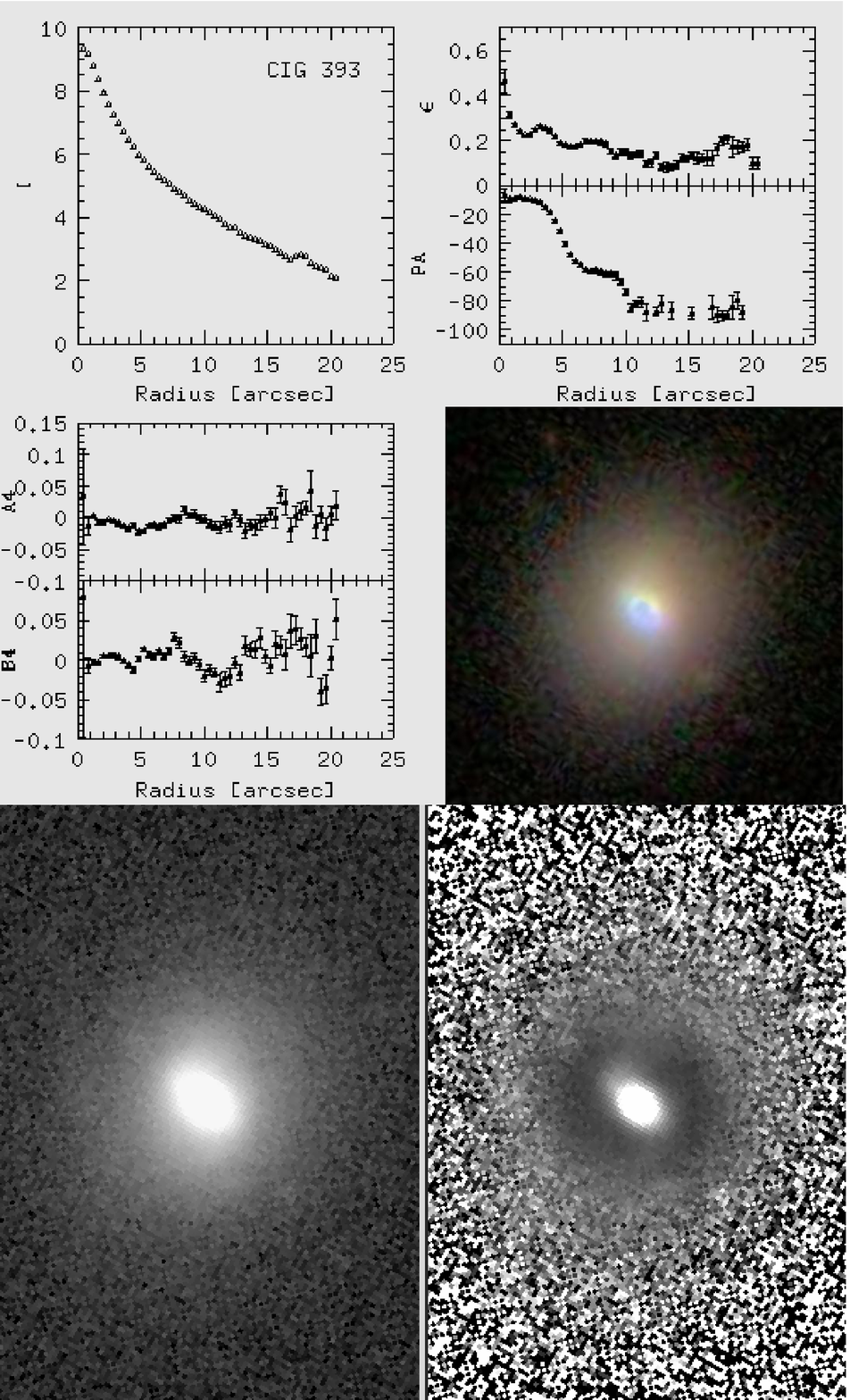}
\caption{CIG 393: a prototype of a very early-type spiral galaxy. Same as in Figure 2.}
\label{mosaico1}
\end{figure*}

Figure 2 shows CIG 722 as a prototype of an E galaxy. The upper panels show the $g$-band surface brightness, $\epsilon$, $PA$ 
radial profiles and $A_{4}/B_{4}$ coefficients of the Fourier series expansions of deviations from pure ellipses. While the surface 
brightness profile shows no trace of a linear component in the outer regions, the $A_{4}$ and $B_{4}$ coefficients of the Fourier series 
expansions do not show any significant tendency towards diskiness or boxiness and at the same time, the $\epsilon$ and $PA$ radial 
profiles show a non-significant trend from the inner to the outer regions. The lower panels show an RGB color map from the SDSS 
database, a logarithmic-scaled $g$-band image and its corresponding filtered-enhanced version. A  uniform color along the face of this 
galaxy is appreciated. The $g$-band filtered-enhanced image shows in addition, some sort of faint external DH's. 
We caution the reader about the existence of these DH's in some of our E/S0 candidates and that care must be taken about its 
interpretation.


Similarly as Figure 2, Figure 3 shows CIG 16 as a prototype of a S0 galaxy. The surface brightness profile clearly shows an inflection 
most probably caused by the underlying disk structure (caution must be paid for the presence of inner rings or bars), the $A_{4}$ and $B_{4}$ 
coefficients of the Fourier series expansions show the tendency to deviate towards disky shapes but this is more clear at the outskirts. At the 
same time, $\epsilon$ indicates a clear change from the inner to the outer regions reaching a maximum and then a more definite value at the 
outer disk region while $PA$ basically indicates the orientation of the semi-major axis. The lower panels present the RGB color map from the 
SDSS database, the logarithmic-scaled $g$-band image, and its filtered-enhanced version. This time, a faint elongated component along the 
main $PA$ of the galaxy can be appreciated in the geometric profiles. The $g$-band filtered-enhanced image let us visualize that disky structure 
suggested in the geometric profiles.

\subsubsection{S0/Sa Types}

Another relevant goal in our study is to isolate as much as possible subtle differences between lenticular S0 and Sa spirals. The advantages 
offered by the  SDSS images allow for a systematic search of fainter features that could point to a more definite morphological classification. 
We observed the combined behavior of both $\epsilon$ and $PA$ radial profiles as an auxiliary tool to disentangle among S0/Sa cases. 
Significant but not necessarily coupled changes in $\epsilon$ and $PA$ radial profiles should be evidencing the presence of additional 
structure (in an already identified disk) in the form of arms, outer rings or envelopes. If further additional image processing did not show 
definite evidence of those features, we kept the galaxy type as lenticular. On the contrary, if the evidence about  features like faint arms are 
definite, we classify the galaxy as Sa.  

Figure 4 illustrates CIG 393 as a prototype of a very early-type Sa galaxy. While originally suspected as a peculiar S0 candidate from its 
appearance in both the $g$-band and RGB SDSS images, the $A_{4}$ and $B_{4}$ coefficients of the Fourier series expansions suggest 
the presence of a disk from the inner to the outskirts and at the same time, $\epsilon$ and $PA$ show a clear non-coupled change resembling 
additional structure(s), likely a set of smooth wide and tightly-wounded arms. The color component  in the central regions appreciated from 
the RGB image also suggests the presence of an extra component. This time the $g$-band filtered-enhanced image allows to see that 
additional component in the form of two inner arms that become wide, smooth but still tight at the outskirts. The oval shape of the bulge 
region suggest a bar signature that $\epsilon$ and $PA$ marginally detect. (c.f.  Wozniak et al. 1995). Cases like CIG 393 clearly illustrates 
the advantages of the higher resolution and depth of the SDSS survey.  


\subsection{Physical Morphology}
\label{PhysMorphology}

Physical morphology has appeared as an alternative for classifying galaxies on the basis of 
measurable structural properties rather than on a visual classification (see Morgan \& Osterbrock 1969; Abraham et al. 1996; Conselice 1997; Bershady, 
Jangren \& Conselice 2000, 
among others). Conselice (2003, and more references therein) has provided a useful framework for classifying galaxies closely tied to 
underlying physical processes and properties. Conselice (2003; hereafter C2003) argues that the major ongoing and past formation 
modes of galaxies can be distinguished using three model--independent structural (photometric) parameters, which allow for a robust 
classification system. These parameters are the concentration of stellar light ($C$), its asymmetric distribution ($A$), and a measure of its 
clumpiness ($S$).

We present our estimate of the $CAS$ parameters in the $r$ band and discuss some trends of the $CAS$ parameters at the SDSS $ugriz$ 
passbands for the retrieved sample of CIG galaxies. The $CAS$ characterization of this CIG galaxies is also helpful as a comparative 
sample for interpreting similar results of other surveys that sample galaxies in similar (Hern\'andez--Toledo et al. 2007; 2008) or wide range 
of environments (e.g., C2003; Hern\'andez--Toledo et al. 2005; 2006). 

We briefly review each one of the $CAS$ parameters.

{\it Concentration of light $C$.-}
The concentration index $C$ is defined as the ratio of the 80\% to 20\% curve of 
growth radii ($r_{80}$, $r_{20}$), within 1.5 times the Petrosian inverted radius 
at r($\eta = 0.2$) ($r_P'$) normalized by a logarithm: 
$C = 5 \times log(r_{80\%}/r_{20\%})$ (see for more details C2003). 
The concentration is related to the galaxy light (or stellar mass) distributions.  

{\it Asymmetry $A$.-}
The asymmetry index is the number computed when a galaxy is rotated $180^{\circ}$ from its center and then subtracted from its pre-rotated 
image, and the summation of the intensities of the absolute value residuals of this subtraction are compared with the original galaxy flux 
(see for more details C2003). This parameter is also measured within $1.5\times r_P'$.  The $A$ index is sensitive to any feature that 
produces asymmetric light distributions. This includes galaxy interactions/mergers, large star-forming regions, and projection effects such as 
dust lanes (Conselice 1997; Conselice et al. 2000). 

{\it Clumpiness $S$.-}
Galaxies undergoing SF are very patchy and contain large amounts of light at high spatial frequency. To quantify this, the clumpiness index 
$S$ is defined as the ratio of the amount of light contained in high frequency structures to the total amount of light in the galaxy within 
$1.5\times r_P'$ (C2003). The $S$ parameter, because of its morphological nature, is sensitive to dust lanes and inclination (C2003).

 {\it Measurement of $CAS$ parameters.-}
The measurement of the $CAS$ parameters for the CIG galaxies was carried out in several steps: 
(i) close field stars were removed from each image; 
(ii) sky background was removed from the images; 
(iii) the center of each galaxy was considered as the barycenter of the light distribution and the starting point for measurements. 

The $CAS$ parameters for  all the S isolated galaxies were estimated directly, i.e. CIG galaxies are not influenced by light 
contamination from any other galaxy of similar size in the neighborhood (isolation criteria). Galaxies with high inclinations or axis 
ratios could introduce systematic biased trends in the values of the $CAS$ parameters (C2003). CIG galaxies whose apparent 
axial ratios yield ``inclinations'' larger than $80^\circ$ are represented as open circles on the corresponding plots. 

To obtain good $CAS$ estimates, the images need to be carefully revised for the possible 
existence of the following problems (that should be eliminated as much as possible): 
(i) images with poor quality or low s/n ratio;
(ii) overlapping bright stars within galaxies;
(iii) galaxies placed at the border of the CCD detector;
(iv) very bright diffraction spikes from nearby stars crossing a galaxy;
(v) galaxies embedded in diffuse light of other sources;
(vi) clear cases of merger candidates.

In the present paper, all those factors have not necessarily been eliminated. Instead a more statistical approach have been assumed 
by restricting the analysis to more definite ranges of $CAS$ values in each band such that values outside those intervals (most likely CIG 
galaxies which images suffer from one or more of the above mentioned drawbacks) were eliminated from the $CAS$ analysis, 
leaving us with a different final number of galaxies in each band. In the $r$ band this number amounts to 390 galaxies.

\begin{figure*}
\plotancho{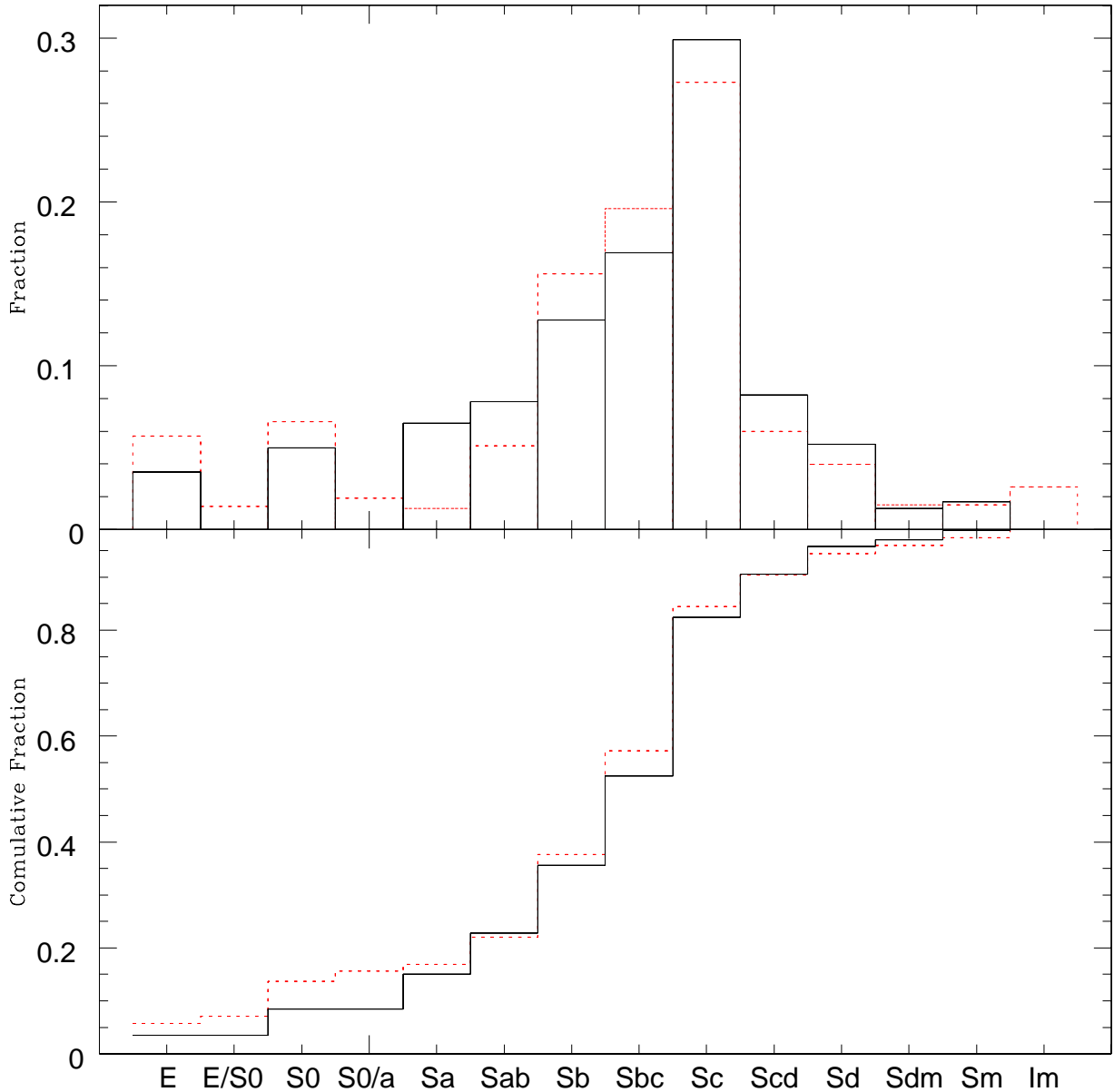}
\caption{Upper panel: Normalized histogram of morphological types for the 539 CIG
galaxies analyzed with our image procedures by using the SDSS (DR6) database (solid line). 
For comparison, the results from Sul06 for a sample of 1018 CIG galaxies classified 
from the POSS II images is also plotted (dashed line).  Lower panel: Cumulative Fraction of 
morphological types found in this work (solid line) and in Sul06 (dashed line). }
\label{mosaico1}
\end{figure*}

\begin{figure}
\plotone{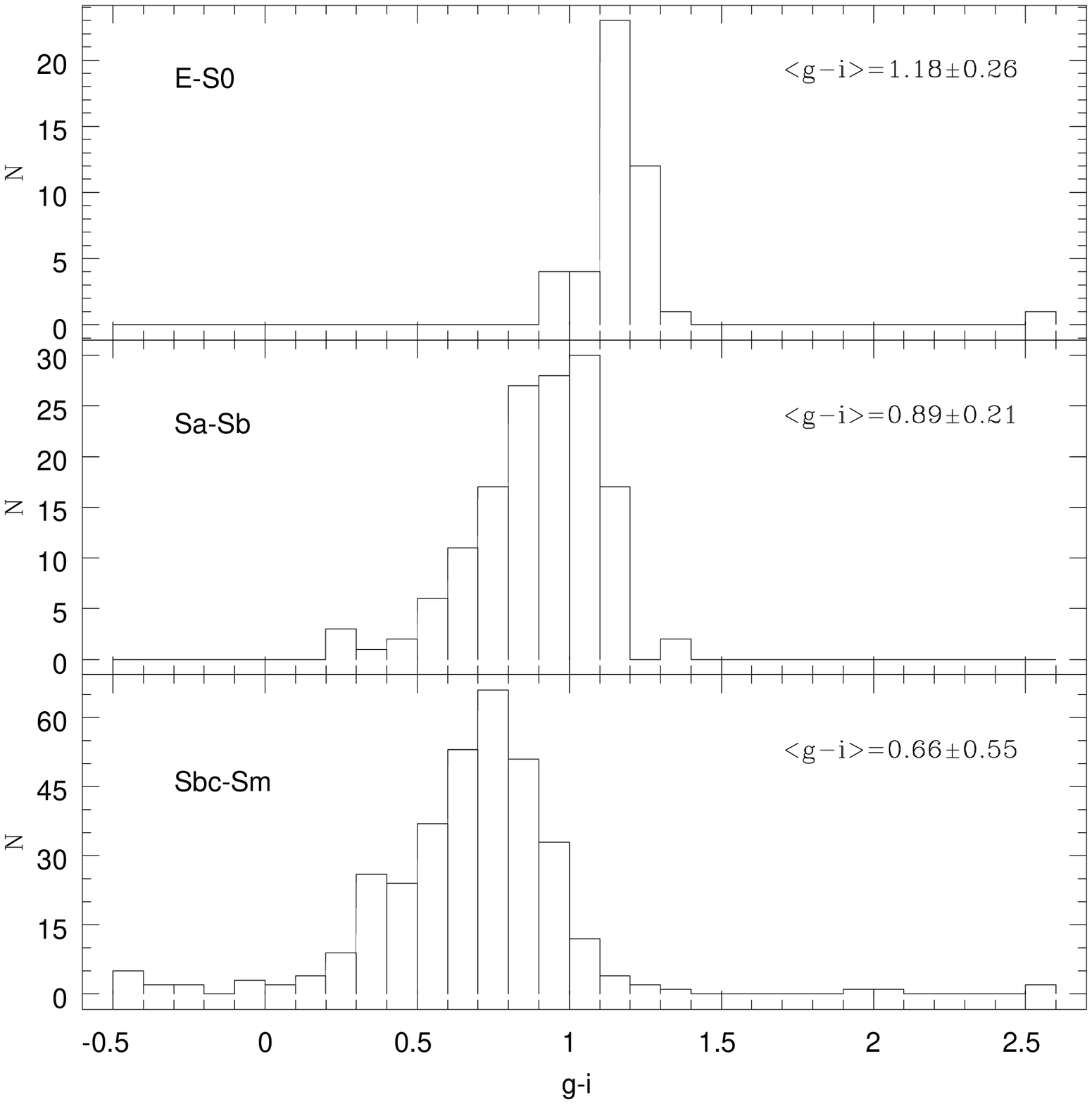}
\caption{$(g-i)$ color histograms for the CIG galaxies sorted into E/S0 types (upper panel), SaSb types 
(middle panel), and SbcSm galaxies (lower panel). }
\label{mosaico1}
\end{figure}

\section{Results}

\subsection{The Atlas}

Each of the 549 isolated CIG$\cap$SDSS galaxies are presented in the form of a mosaic. 
For S galaxies of types later than Sa, we include, from upper-left 
to lower-right panel:
(1)  a gray scale $g$-band image displayed in logarithmic scale ,
(2) an $g$ band filtered-enhanced version of the image in 1), and
(3) an RGB color image from the SDSS database (See Figure 1).

For E/S0/Sa galaxies, besides including the processed  images  1), 2) and 3) above, additional surface brightness profiles and  geometric 
profiles (ellipticity $\epsilon$, Position Angle $PA$ and $A_{4}/B_{4}$ coefficients of the Fourier series expansions of deviations of a pure 
ellipse) from the $r$ band images are provided (See Figures 2, 3 and 4).

All the 549 mosaic images in the present revaluation can be reviewed on the web-site 
(http://132.248.1.210).

\subsection{Optical Morphology}
   
Table 1 reports our morphological revaluation of 539 CIG galaxies after eliminating 10 galaxies in common to the lists of possible interacting 
galaxies by Sul06. Columns (1), (2) and (3) give the morphological type, the number $n$ of galaxies in each morphological type and the 
corresponding fraction $n/1018$ as reported in Table 2 of Sul06. Similarly, Columns (4) and (5) give the number $n_{\rm SDSS}$ of retrieved 
galaxies in this work and the estimated fraction $n_{\rm SDSS/539}$ for each morphological type according to our revaluation.

\placetable{tbl-1} 
\begin{deluxetable}{lccccccc} 
\tablecolumns{5}
\tablewidth{0pc} 
\tablecaption{Results of the new morphological classification for the  CIG$\cap$SDSS sample based on 
the available SDSS (DR6) data.
\label{tbl-1}}
\tablehead{
\colhead{} Type & $n$ & $n/1018$ & $n_{SDSS}$ & $n_{SDSS}/539$ &}  
\startdata 

E       & 58    & 0.057 & 19   & 0.035 &  \\
E/S0 & 14    & 0.014 &  -      &    -       &  \\
S0     & 67    & 0.066 & 27   & 0.050 &  \\
S0/a  & 19    & 0.019 &    -    &       -   &  \\
Sa     & 13    & 0.013 & 35   & 0.065 &  \\
Sab   & 52    & 0.051 & 42   & 0.078 &  \\
Sb     & 159  & 0.156 & 69   & 0.128 &  \\
Sbc   & 200  & 0.196 & 91   & 0.169 &  \\
Sc     & 278  & 0.273 & 161 & 0.299 &  \\
Scd   & 61    & 0.060 & 44   & 0.082 &  \\
Sd     & 41    & 0.040 & 28   & 0.052 &  \\
Sdm  & 15    & 0.015 & 7     & 0.013 &  \\
Sm    & 15    & 0.015 & 9     & 0.017 &  \\
Im      & 26   & 0.026 & 5     & 0.009 &  \\
\hline 
E-S0  & 125   & 0.122 & 46   & 0.085 & \\
Sa-Sd & 804 & 0.790 & 470 & 0.872 & \\
Sb-Sc & 637 & 0.626 & 321 & 0.596 & \\

\enddata
\end{deluxetable}


Figure 5 shows the differential and cumulative histograms of the results in Table 1. 
Solid lines correspond to the reevaluated fractions in this work.
We find that the median type of the CIG$\cap$SDSS sample is Sbc, with 35\% of the 
galaxies being earlier than Sbc, and 65\% being Sbc type or latter. A fraction of  90.20\% are 
S galaxies in the range of Sa--Sm types and only 8.53\% are of early--type (E--S0); the 
remaining 1.27\% are irregulars (0.9\%) and unclasificable galaxies (0.37\%, which corresponds
to two objects). The fraction of early-type spirals (Sa and Sab) amounts to about 14\%.  
These results are at odds with those by Sul06 (dashed line in Fig. 5), who 
classified 14\% of the CIG galaxies as E/S0 types and 82\% as Sa-Sd types. Among the 
spirals, only 6\% were classified as early--type Sa--Sab spirals. Notice that we attempted 
for a cleaner classification avoiding transition E/S0 and S0/Sa cases.  A suitable processing  
of the SDSS images complemented with color information permit a better spatial discrimination 
of morphological details in spiral galaxies like the prominence and extent of bulges. Higher 
resolution is important to distinguish between inner rings and ring-like features produced 
by the tightening of the arms, the fragmentation degree of the arms or structures like bars, 
clumps, dust lanes, among others. Our filtering process enhances high spatial frequency 
structures in S's and (depending on the kernel size) in E's too.  

Our preliminary bar identification is based on a careful visual inspection of the 
$r-$band clean/subtracted images as well as the corresponding filtered--enhanced and 
RGB images. For early--type galaxies, the surface brightness and geometric profiles, after 
an isophotal analysis, were also used to define the presence of a bar. As the result,
we identified 292 disk galaxies with bars, which accounts for 59.2\% of the disk galaxies 
in the CIG$\cap$SDSS sample (25.5\% with clear bars, SB type, and 33.7\% with weak or 
suspected  bars, SAB type). Because bar detection becomes difficult in highly inclined
galaxies, we adopted also the standard procedure of excluding objects with an inclination
$i\ge 60^{\circ}$ ($\approx 56\%$ of the disk galaxy subsample). For the $i\le 60^{\circ}$
sub--sample, we find a fraction of barred galaxies of 65.8\% (SA=35\% and SAB=30.8\%). 
The bar fraction in early type spirals (Sa--Sb) is 65\% 
and is almost the same one (66\%) in late types (Sbc--Sm/Irr), though we caution the reader 
about possible biases in this result (see \S\S 5.2 for a discussion).
A more careful analysis including more appropriate ways to detect bars for this sample will be 
presented elsewhere. Similarly, the information for rings in disk galaxies 
(inner, outer rings and pseudo--rings) in our sample is tentatively available for 164
galaxies accounting to 33.3\% of the disk galaxies.

Table 2  shows a comparison of our morphological results against the corresponding information from the literature. Only 
CIG galaxies with $V_{rad} > 1000\, km s^{-1}$ are included. Column (1) is the CIG cataloged number, Column (2) the Hubble Type 
from Lyon Extragalactic Database (HyperLeda), Column (3) the Hubble Type from NED, Column (4) the Hubble Type from Sul06 Column 
(5) the Hubble Type from this work, and finally  Column (6) the inclination to the line-of-sight as reported in the HyperLeda database.

Table 3 shows similar results as Table 2 but for CIG galaxies with ($V_{r} < 1000\, km s^{-1}$).  Although at $V_{r} < 1000\, km s^{-1}$ the 
morphology is more difficult to evaluate in terms of isolation, we have proceeded because in nearby galaxies more detailed structural 
information is generally available, thus providing another test on the reliability of our classification.  The inclination is included as a guide 
to evaluate the reliability of the classification.

\placetable{tbl-2}
\begin{deluxetable}{cccccccc} 
\tablecolumns{8}
\tablewidth{0pc} 
\tablecaption{Detailed morphology for the CIG$\cap$SDSS sample ($V_{r} > 1000\, km s^{-1}$).
\label{tbl-1}}
\tablehead{
\colhead{} CIG    &    Leda    &    NED    &    Sulentic 2006    &    This work    &    incl     &   }
\startdata 

CIG 0011    &    SABc    &    SAB(s)c?    &    Sbc    &    SABc    &    67.48    &    \\
CIG 0012    &    Sbc    &    Sb    &    Sb    &    Sbc    &    79.45    &   \\
CIG 0016    &    Sbc    &    S0    &    S0    &    S0    &    64.22    &     \\
CIG 0019    &    S0-a    &    S0    &    S0    &    SABa    &    90.00    &    \\
CIG 0033    &    SABc    &    SAB(rs)cd    &    Sb    &    SABc    &    53.76    &    \\
CIG 0056    &    Sb    &    SB(rs)b    &    Sb    &    RSBb(r)    &    52.52    &    \\
CIG 0060    &    Sb    &    Sb    &    Sd    &    SABc    &    62.87    &    \\
CIG 0081    &    Sb    &    Sb    &    S0    &    S0    &    84.80    &     \\
CIG 0187    &    SABb    &    SAB(s)bc    &    Sc    &    SBbc    &    23.76    &   \\
CIG 0189    &    E    &    Sa    &    E    &    E    &    90.00    &     \\
:    &       &       &       &       &       &       & \\

\enddata
\end{deluxetable}

\placetable{tbl-3}
\begin{deluxetable}{cccccccc} 
\tablecolumns{8}
\tablewidth{0pc} 
\tablecaption{Detailed morphology for the CIG$\cap$SDSS sample ($V_{r} < 1000\, km s^{-1}$).
\label{tbl-1}}
\tablehead{
\colhead{} CIG    &    Leda    &    NED   &   This work    &          incl           &     }  
\startdata 

CIG 0190     &     IB     &     Im     &               Irr     &     62.62     &            \\
CIG 0193     &     Scd     &     SA(s)d     &               SABcd     &     61.27     &              \\
CIG 0224     &     Scd     &     SB(rs)d     &               SBd     &     20.61     &               \\
CIG 0235     &     SBd     &     SB(s)mpec     &               Sd     &     68.71     &           \\
CIG 0265     &     Sbc     &     SABdm     &               SABdm     &     62.44     &              \\
CIG 0324     &     Sb     &     SA(r)b     &               SABa     &     68     &              \\
CIG 0347     &     SABb     &     SB(s)d     &         SBc     &     56     &               \\
CIG 0388     &     IB     &     Im     &               Irr     &     59.88     &             \\
CIG 0428     &     SBc     &     SB(rs)cd     &               SBcd     &     62.69     &           \\
CIG 0434     &     I     &     Im     &               Sm     &     24.59     &            \\
:     &         &        &          &          &          \\

\enddata
\end{deluxetable}

\subsection{$CAS$ Results}

The $CAS$ parameters in the recent literature are frequently reported in the Johnson-Cousins $R$ band. Given the similarity in definition 
(both in $\lambda_c$ and $\Delta \lambda$) between $R$ and the SDSS $r$ band photometric systems, a similarity in the estimated $CAS$ 
values is expected. Thus we report here the $CAS$ parameters and the corresponding errors in the $r$ band for a sample of 390 CIG 
galaxies. The $CAS$ parameters are listed in Table 4. The $CAS$ values for the CIG galaxies in other bands can be provided by the 
authors upon request.

We have sorted the CIG sample into three groups: early-- and late--type spirals (SaSb and SbcSm, respectively)  and E/S0 galaxies. The 
corresponding average and standard deviation values of the $CAS$ parameters are listed in Table 5. 
 
In spite of the different zero-point scales that define the SDSS $r$ band and the Johnson-Cousins $R$ band systems, we find similar 
values as those reported in the $R$ band by Conselice (2003) for the Frei et al. (1996) sample of non--interacting galaxies and also for 
smaller subsamples of CIG galaxies in Hern\'andez-Toledo et al. (2007; 2008).

 \begin{figure}
\plotone{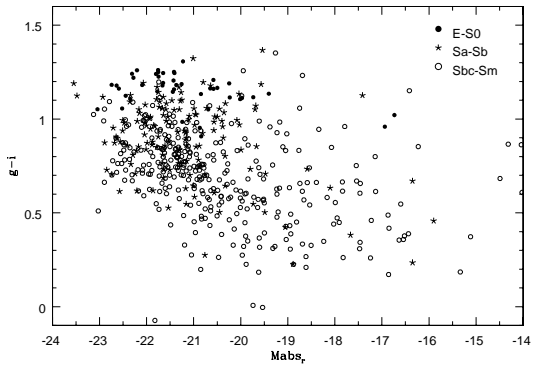}
\caption{ Color-Magnitude ($g-i$ vs $M_{r}$) diagram for the CIG galaxies sorted into E/S0 types (solid circles), SaSb types 
(asterisks), and SbcSm galaxies (open circles). }
\label{mosaico1}
\end{figure} 

\subsection{General properties}

We have calculated the absolute magnitudes in {\it g, r,} and $i$ bands for our sample
of 539 CIG$\cap$SDSS galaxies by using the HyperLeda distance modulus with $H_0=70\, kms^{-1}Mpc^{-1}$.
We have used the apparent magnitudes corrected by Galaxy extinction ($A_{\lambda}^b$[mag]) 
given for each galaxy in the SDSS database; the corrections are based on the Schlegel et al. 
(1998) Galaxy maps.  For the 493 disk galaxies in our sample, we have further corrected the
magnitudes by internal extinction. Several pieces of evidence suggest that the fraction of dust 
is larger for bigger galaxies. Therefore, the internal extinction correction should depend not 
only on inclination but also on galaxy scale: $A_{\lambda}^i$[mag] = $\gamma_\lambda$log($a/b$), 
where $a/b$ is the major--to--minor axis ratio (taken from HyperLeda), and 
$\gamma_\lambda$ is a scale--dependent coefficient in the given passband $\lambda$.  
From an empirical analysis, Tully et al. (1998) inferred the coefficients $\gamma_\lambda$ 
in the $BRIK$ bands as a function of the galaxy maximum circular velocity. From their data 
(reported explicitly in Tully \& Pierce 2000), Hern\'andez--Toledo et al. (2007) have carried 
out linear correlations of these calculated coefficients with the corresponding magnitudes 
{\it not corrected} for internal extinction, $M_{\lambda}^b$. Here, we interpolate these 
linear correlations to the SDSS $gri$ bands and obtain the following approximate coefficients:
\begin{eqnarray}
\gamma_g {\rm [mag]}= -6.00 -0.38 M_g^b, \cr 
\gamma_r {\rm [mag]}= -4.00 -0.25 M_r^b, \cr 
\gamma_i {\rm [mag]}= -3.60 -0.21 M_i^b. 
\label{extin}
\end{eqnarray}
These coefficients are not valid for low--luminosity galaxies (see Hern\'andez--Toledo 
et al. 2007 for the lower limits in the $BVRIK$ bands); for the few cases
when $\gamma_\lambda<0$, we set $\gamma_\lambda=0$. We stress that the internal extinction 
correction suggested here is of statistical nature and it provides only an approximation 
to a complicated problem, however, it is certainly better than not 
correcting the data at all for internal extinction. The final corrected
magnitude for disk galaxies is then 
$M_{\lambda}^{b,i} = M_{\lambda} - A_{\lambda}^b - A_{\lambda}^i$.

\placetable{tbl-4}
\begin{deluxetable}{cccccccc} 
\tablecolumns{7}
\tablewidth{0pc} 
\tablecaption{$CAS$ structural parameters and their errors in the $r$ band for the CIG$\cap$SDSS sample.
\label{tbl-1}}
\tablehead{
\colhead{}  CIG    &    C   &   errC &    A   &   errA &    S   &   errS & }  
\startdata 

CIG0011   &   3.363   &   0.084   &   0.176   &   0.061   &   0.060   &   0.013    &   \\
CIG0012   &   3.169   &   0.166   &   0.302   &   0.040   &   0.230   &   0.020    &   \\
CIG0016   &   3.603   &   0.371   &   0.095   &   0.007   &   0.190   &   0.008    &   \\
CIG0019   &   3.686   &   0.339   &   0.103   &   0.006   &   0.120   &   0.005    &   \\
CIG0033   &   2.962   &   0.105   &   0.190   &   0.017   &   0.380   &   0.018    &   \\
CIG0056   &   4.193   &   0.128   &   0.174   &   0.020   &  -0.030   &   0.004    &   \\
CIG0060   &   2.723   &   0.189   &   0.225   &   0.020   &   0.320   &   0.018    &   \\
CIG0081   &   4.107   &   0.249   &   0.085   &   0.005   &   0.020   &   0.001    &   \\
CIG0187   &   3.224   &   0.072   &   0.207   &   0.066   &   0.600   &   0.043    &   \\
CIG0189   &   4.446   &   0.174   &   0.128   &   0.007   &   0.280   &   0.009    &   \\
:   &      &      &     &      &      &      &   \\

\enddata
\end{deluxetable}
\placetable{tbl-5}
\begin{deluxetable}{ccccccc} 
\tablecolumns{8}
\tablewidth{0pc} 
\tablecaption{Mean $r$ band $CAS$ parameters and the corresponding 1 $\sigma$ for CIG$\cap$SDSS
galaxies according to morphological types.\label{tbl-2}}
\tablehead{
\colhead{} $CAS$ & E/S0 & SaSb & SbcSm & }  
\startdata 

C  & 3.81 $\pm$ 0.46 & 3.58 $\pm$ 0.52 & 2.97 $\pm$ 0.49 & \\
A  & 0.09 $\pm$ 0.03 & 0.14 $\pm$ 0.06 & 0.20 $\pm$ 0.07 & \\
S  & 0.08 $\pm$ 0.10 & 0.16 $\pm$ 0.16 & 0.25 $\pm$ 0.18 & \\
\enddata
\end{deluxetable}

By using the corrected absolute magnitudes, we are able to calculate the galaxy corrected
colors. In Fig. 6 the $g-i$ color histograms of the 539 CIG$\cap$SDSS galaxies  
separated into three morphological ranges are plotted. The averages and 1$\sigma$ dispersions
of $g-i$ for the E/S0, Sa--Sb, and Sbc--Sm/Irr morphology ranges are
$1.18\pm 0.26$, $0.89\pm 0.21$, and $0.66\pm 0.55$, respectively.  Notice that in the case of E/S0 galaxies 
no internal extinction correction was applied. 


Figure 7 shows the color--magnitude
diagram for all the sample, again separated into three morphological ranges (plotted with 
different symbols). The E/S0 galaxies occupy a quite flat and narrow region in the diagram, 
while for disk galaxies a loose trend of redder colors as their are brighter is seen, but with a 
large scatter.  The average $r$-band absolute magnitudes are $<-20.91 \pm 2.21>$, $<-21.02 \pm 1.61>$,
and $<-19.84 \pm 2.84>$, for E/S0, SaSb and galaxies later than Sbc types, respectively. From these results, we remark that
(i) in isolated environments, there is a paucity of luminous E/S0 galaxies,  and (ii) 
the average luminosity of E/S0 galaxies is the same as that in early-type spirals and slightly more luminous than late-type spirals. 
Park et al. (1994) first quantified the paucity of luminous galaxies at very low environments, they studied 
the distribution of local density around galaxies showing that regions of moderate and high density contain both very bright 
(M $\leq M_{*}$ = -19.2 + 5 log h) and fainter galaxies, but that voids preferentially harbor fainter galaxies (approximately 2$\sigma$ 
significance level). This last result was later confirmed by Hoyle et al. (2005) by studying the luminosity function of void galaxies in the 
SDSS.

Notice that our results are in contrast to those in Sul06  where late-type spirals were found to be more luminous on average than E/S0 
galaxies.  Park et al. (2007) examined the color-magnitude ($u-r$ vs $M_{r}$) diagram for early and late-type SDSS 
galaxies located in high and low density environments, finding a very slight dependence of the location of the early-type SDSS 
galaxies on the density field and a stronger dependence of the late-type colors on environment.


\section{Discussion}

\subsection{Optical morphology: comparison with previous works}

Figure 8 shows a set of histograms comparing the results of our morphological revaluation against those in the literature.  The upper 
left panel compares our morphological types vs  HyperLeda ($\Delta T = T_{Ours} - T_{Leda}$). The upper-right panel  compares our 
morphological types vs  NED ($\Delta T = T_{Ours} - T_{NED}$) and finally, the lower-left panel is a comparison of our results vs those 
from the morphological revaluation by Sul06 ($\Delta T = T_{Ours} - T_{Sul06}$). The type code number from RC3 and the additional 
convention suggested for early types in Sul06 were adopted. 


Out from 488 galaxies in common with HyperLeda, 29\% have $\Delta T = 0$ and  up to 64\% have a difference $|\Delta T | < 2$. One 
third more show larger morphological differences that appear biased towards small positive differences. Similarly, from 496 galaxies in 
common with NED, 22\% have $\Delta T = 0$ and up to 49\% of the sample have $|\Delta T | < 2$. This time a larger fraction from the 
complementary 50\% are distributed towards small positive differences. This suggests that in spite of the large  dispersion ($\Delta T = 10$) 
shown in both plots, HyperLeda appears to contain more reliable morphological information previous to the POSS II revaluation from Sul06. 
The lower panel of Figure 8 shows that from 502 galaxies in common with Sul06, 40\% have $\Delta T = 0$ and up to 74\% of them have 
$|\Delta T | < 2$, indicating a closer agreement in the derived types. The dispersion has significantly decreased ($\Delta T = 5$) and only 
25\% of the sample is distributed evenly in this range. Only a slight difference is still apreciated first in the negative $\Delta T = -2$ side and 
then in the positive $\Delta T > 2$ side, emphasizing the ability of the SDSS data and the corresponding processing to resolve finer 
morphological details either in early and late-type galaxies.

Figure 9 shows similar results as those in Figure 8 but for CIG galaxies with $V_{r} < 1000\, km s^{-1}$. This time the histograms compare 
of our results against those in HyperLeda (upper left panel), NED (upper-right panel) and other authors as compiled from Sul06 (lower-left 
panel). When more than one possible classification for each CIG galaxy is available in the compilation of Sul06, we adopted the type closer to 
our results. 

\begin{figure}
\plotone{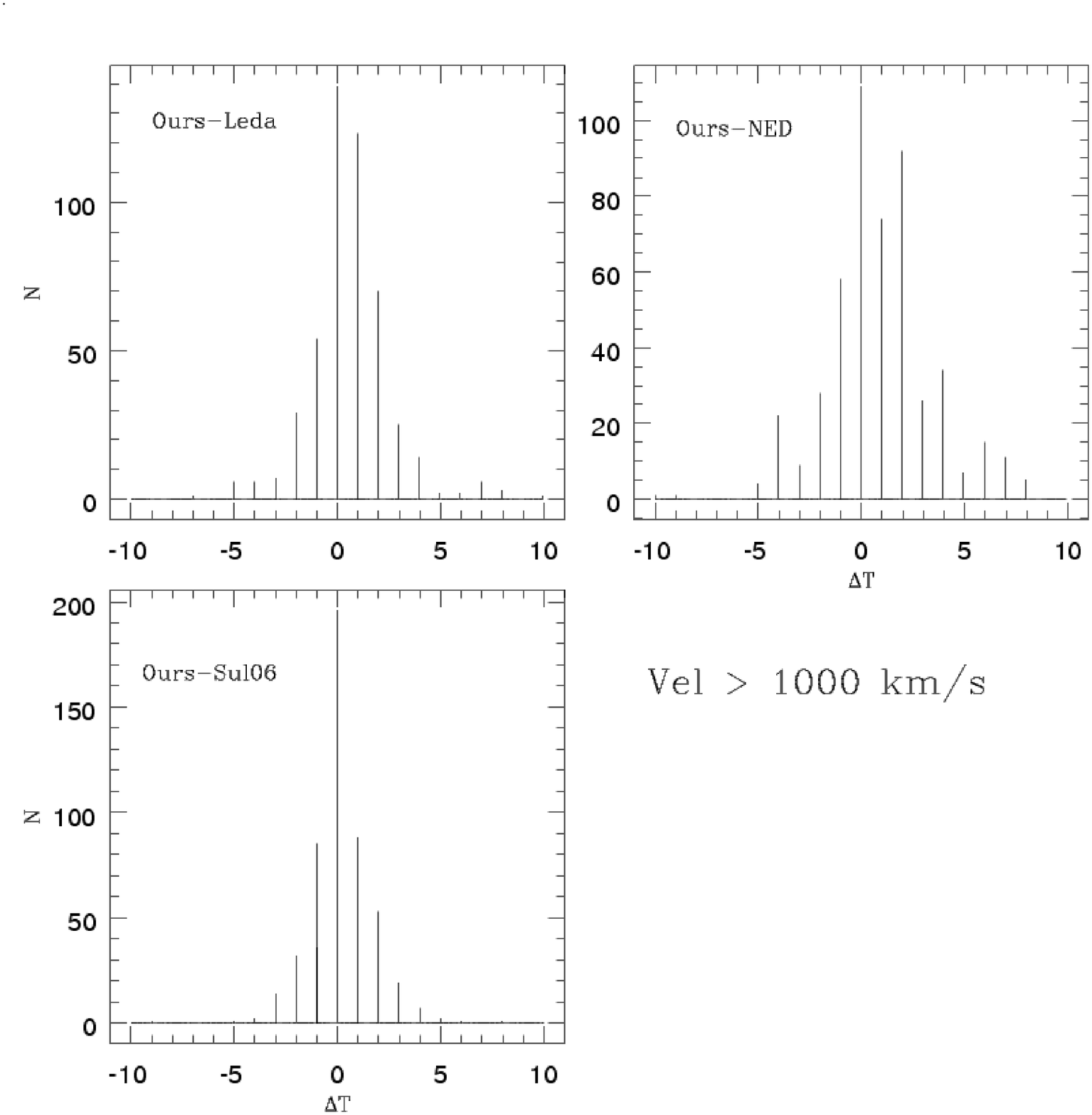}
\caption{Histograms of morphological type differences $\Delta T = T_{ours}-T_{other}$ for galaxies with $V > 1000$, 
where $T_{ours}$ is the reevaluated morphological type in this work, and $T_{other}$ is the 
morphological type reported in LEDA (upper left panel), NED (upper left panel), and Sul06 
(lower panel). }
\label{mosaico1}
\end{figure}

\begin{figure}
\plotone{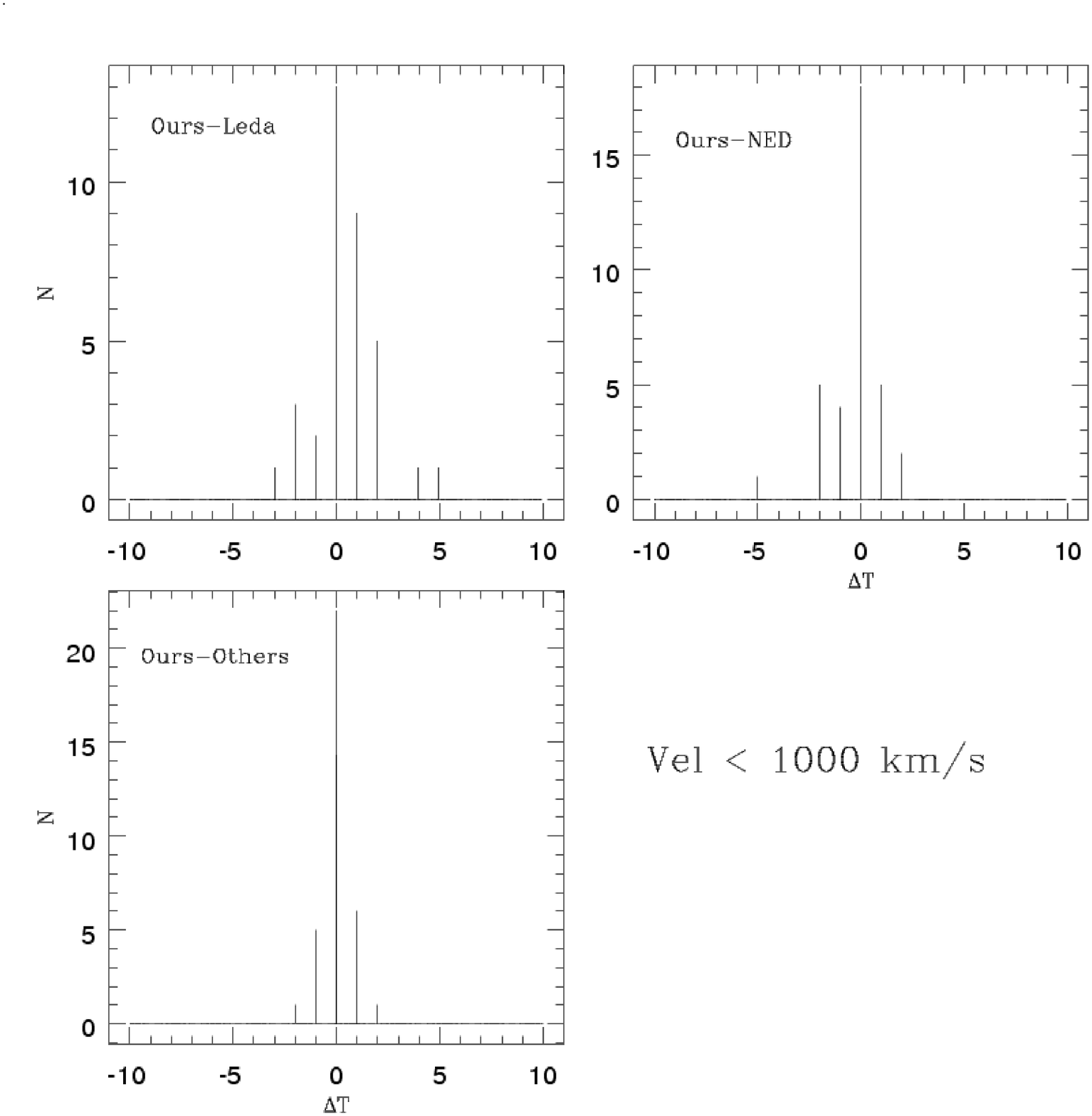}
\caption{Histograms of morphological type differences $\Delta T = T_{ours}-T_{other}$ for galaxies with $V < 1000$.
Upper left panel: this work - Leda, right panel: this work - NED and lower panel: this work - other sources. 
Other sources include individual observations of diverse nature (photographic, photoelectric and CCD) form different authors.  }
\label{mosaico1}
\end{figure} 

Out from 35 galaxies in common with HyperLeda, 37\% have $\Delta T = 0$ and up to 69\% have $|\Delta T | < 2$. Still one third more show 
slightly larger morphological differences that appear biased towards small positive differences. This reflects the uniformity in the classification 
data provided in HyperLeda. However, from 35 galaxies in common with NED, 51\% have $\Delta T = 0$ and up to 77\% of the sample have 
$|\Delta T | < 2$. The scatter have significantly narrowed, and this time NED appears to contain more detailed morphological information for 
nearby objects, contrary to the situation for $V_{r} > 1000\, km s^{-1}$, probably suggesting a more inhomogeneous nature of the compiled data 
in that database. Finally, the lower panel of Figure 9 shows that from 35 galaxies in common with other authors, 63\% have $\Delta T = 0$ and 
up to 94\% of them have $|\Delta T | < 2$, indicating a significant agreement with our derived types. This is not surprising since nearby 
galaxies are subject to a more precise morphological classification either from high quality photographic or CCD data.


\subsection{Bars in isolated S galaxies}

Our images show that bars in CIG galaxies come into a variety of sizes, shapes and color 
distributions; from apparently strong to small ones confined to the central parts of galaxies 
and up to the oval-shaped bulges, suggesting a range of strengths, lengths, and mass distributions.
 The tentative fraction of bars found here for the isolated S galaxies
with inclination $i\le 60^{\circ}$ is 66.8\% (SB=35\% and SAB=30.8\%); by relaxing the inclination
condition, the resultant fraction is about 59\%.

As Barazza, Jogee \& Marinova (2008) remark, the analyzed galaxy samples in previous works
used to be small in number and dominated by early--type spirals. Besides, these samples are for
galaxies in any environment, mostly from the field, but not selected to be isolated. The
reported fractions of barred disk galaxies in the literature (in the optical) amounts to 
$\sim 45\%$, while in the near--infrared, the fractions seem to increase up to $\sim 60\%$ 
(see e.g., Eskridge et al. 2000; Marinova \& Jogee 2007, and more references therein). For 
the large sample of local SDSS disk galaxies analyzed recently by Barazza et al. (2008), 
the fraction of 
{\it large--scale} bars in the $r$ band is $\sim 48\%-52\%$. This fraction is smaller 
than the one reported here. However, we should have in mind that (1) in Barazza et al. 2008
only large--scale bars were considered, a condition that we
did not impose, and (2) our bar analysis is based on visual inspections rather than on 
quantitative measurements; only a small fraction of spirals (mainly Sa types) were 
properly judged for the presence of a bar through a deeper analysis of the isophotal 
ellipticity and $PA$ profiles. Thus, it might be that our bar fraction in the $r-$band is 
overestimated with respect to Barazza et al. On the other hand, we emphasize that the observed 
fraction of bars (and rings) in the present paper can hardly be a bias of object selection 
since we simply gathered galaxies according to their availability on the SDSS database.

If confirmed our finding of about a 65\% fraction of bars in the isolated S galaxies ($r$ band), 
it would imply that interactions and the global effects of the dense environment are not crucial 
for the formation of bars, if any, the opposite applies, i.e. the bars tend to be
thickened or destroyed more efficiently in the higher density media. The bar thickening
or destruction is thought to be at the origin of the boxy/peanut bulges or pseudobulges, 
and recent simulations in the context of 
the $\Lambda$CDM cosmology show the viability of this proposal (e.g., Avila-Reese et al. 
2005; Athanassoula 2005; Bornaud, Combes \& Semelin 2005;  Martinez-Valpuesta, Shlosman 
\& Heller 2006). It could be that in the bar phenomenon dominate the internal disk galaxy 
processes rather than the
environmental influence (see for a related discussion Hern\'andez-Toledo et al. 2007).

As in previous works, we also found here that the bar occurrence is similar
in early-- and late--type spirals (c.f. Eskridge et al. 2000; Hern\'andez-Toledo et al. 2007
but see Barazza et al. 2008). What the latter authors report actually is a clear tendency 
to increase the bar fraction as the $r_e/R_{24}$ ratio increases ($r_e$ and $R_{24}$ are 
the effective and optical radius, respectively, and their ratio can be interpreted as 
a measure of the light concentration, and therefore of the bulge--to--disk ratio). For our 
sample, we have measured the $C_{80/20}$ concentration, and we confirm that indeed the
fraction of barred galaxies significantly depends on it: for
$C_{80/20}\le 2.5$, $2.5<C_{80/20}<3.5$, and $C_{80/20}\geqslant 3.5$ the fraction of barred
galaxies increases from $\approx 55\%$ to $\approx 63\%$ and $\approx 78\%$, respectively. 
Therefore, we can conclude that while the fraction of barred galaxies does not 
depend on the morphological type, it does as a function of the light concentration parameter.   
It is definitively crucial to carry out a deeper study to infer the fraction of bars in 
samples like the CIG one by using reliable and homogeneously applied bar--detecting techniques.
 
\begin{deluxetable*}{llllllllll} 
\tablecolumns{10}
\tablecaption{Isophotal Shape, Fine Structure, Dust Lanes and other internal features in Isolated Ellipticals. 
$DH$ remains for a Diffuse Halo component observed in some isolated Ellipticals. \label{tbl-1}}
\tablehead{
\colhead{} CIG  &  boxy/disky  &  fine   & dust & inner & knots &$DH$ & $g-i$ & $M_{r}$ \\      
                &              & structure & lanes & disks &   &    &       &    }  
\startdata 

 CIG0189  &       boxy   &                        &           & disk &  knot  &    &  1.16  &  -20.58 \\
 CIG0245  &              &  shell/ring            &           & disk &        &$DH$&  1.11  &  -19.73 \\
 CIG0264  &      boxy    &                        &           & disk &  knot  &$DH$&  0.98  &  -21.24 \\
 CIG0305  &              &  shell?                &           &      &        &$DH$&  1.16  &  -20.64 \\
 CIG0396  &              &                        &           &      &        &    &  1.11  &  -19.97 \\
 CIG0437  &      disky   &                        &           &      &        &    &  1.26  &  -21.76 \\
 CIG0513  &      boxy    &  ripple?               &           &      &  knot  &$DH$&  1.14  &  -21.77 \\
 CIG0517  &      boxy    &                        &           &      &  knot  &$DH$&  1.19  &   -21.65 \\
 CIG0533  &      boxy    &  ripple?               &           &      &        &$DH$&  1.18  &   -22.74 \\
 CIG0556  &      disky   &  ripple?               &           &      &        &    &  1.02  &  -16.73 \\
 CIG0557  &      disky   &  shell                 &           & disk &        &    &  1.24  &  -22.31 \\
 CIG0578  &      disky   &                        &           &      &        &$DH$&  1.20  &  -21.43 \\
 CIG0582  &      boxy    &                        & dust      &      &  knot  &    & 1.25   &  -21.42 \\
 CIG0599  &      disky   & shell/ring             &           &      &        &    & 1.12   &  -22.44 \\
 CIG0705  &      disky   & shell/ring             &           & disk &        &    & 1.1    & -22.35 \\
 CIG0722  &      boxy    &                        &           &      &        &    & 1.18   & -22.64 \\
 CIG0732  &      boxy    &                        & dust      & disk?&        &    & 1.18   & -22.01 \\
 CIG0768  &      disky   & shell/ring             & dust      & disk &        &    & 0.91   & -20.84 \\
\enddata
\end{deluxetable*}

\subsection{Morphological distortions in the isolated E galaxies}

The morphological properties of nearby E and S0 galaxies in different environments reveal 
important clues for understanding the nature of their formation. The detection of fine 
structures, dust lanes, blue cores, and nuclear disks in early-type galaxies, considered 
evidence of recent merging/accretion events (e.g., Malin \& Carter 1983; Lauer 1985; Abraham 
et al. 1999; Menanteau et al. 2001; Papovich et al. 2003; Lauer et al. 2005; van Dokkum 2005;
Bell et al. 2006), tend to occur more 
frequently in galaxies in the field than in cluster members (Schweizer 1992; Reduzzi et al. 1996; 
Kuntschner et al. 2002). On the other hand, while the bulk of the stars in luminous cluster 
E galaxies are old ($z>2$) and coeval (e.g., Bender et al. 1997; Heavens et al. 2004), in 
low-density environments the colors and color gradients of E galaxies suggest that recent bursts 
of SF occurred (Menanteau et al. 2001, 2005; Stanford et al. 2004; Treu et al. 2005). Blue clumps 
in early-type galaxies have also been suggested as being evidence of recent accretion episodes 
(Elmegreen et al. 2005; Pasquali et al. 2006). According to these authors,  the fraction of 
early-type galaxies with blue clumps increases at high redshift.

A long--standing problem of galaxy formation models is related to the existence and the
properties of E galaxies in the field. Within the context of the hierarchical CDM model, 
the natural prediction is that the stars in field luminous E galaxies should be younger on average 
than those in cluster luminous E galaxies (by $\approx 4$ Gyr, according to models by Kauffmann 
1996). In other words, an isolated luminous E galaxy is expected to assemble late, while
its cluster counterpart, assembles early in an overdense region and then its evolution is
frozen after this region becomes a dynamically relaxed system (the probability of mergers
falls drastically and the mass accretion is even reversed due to tidal stripping and 
intracluster medium ram pressure). Observational studies show that the
stellar populations of field luminous E galaxies are indeed younger than those of cluster
E's. However, according to most of the studies, the age differences are small (e.g., 
Treu et al. 2005; Bernardi et al. 2006; van Dokkum \& van der Marel 2007) or intermediate 
(up to 2 Gyr, e.g., Thomas et al. 2005; Clemens et al. 2006), but not as large as 4 Gyr. 
Besides, according to Treu et al. (2005), the SF history in the field E's 
depends strongly on luminosity: the most luminous E's have old stellar populations, while
the less luminous ones, show evidence of a significant fraction of stellar mass formed
relatively recently.  The color-magnitude diagrams as a function of type and local density 
in Park et al. (2007) show a slight shift of the early-type sequence toward blue color when the 
local density changes by about an order of magnitude, consistent with the above statements.

The tension between models and observations is reduced significantly when possible effects 
of AGN feedback are introduced (de Lucia et al. 2006). Nevertheless, still remains the question 
whether the ``dynamical'' assembling of field E's was recent or not. The de Lucia et al. (2006) 
model predicts that massive isolated E's assembled typically 50\% of their final mass at $z<0.8$; 
such a process is expected to be driven mainly by dry mergers: major mergers between early--type
galaxies without the presence of gas. From the observational point of
view, it was not an easy task to find pieces of evidence of dry mergers since they progress
rapidly, and (i) the post--merger distortions tend to disperse quickly and (ii) they
are not in general of easy detection.  Massive and detailed morphological studies of 
early--type galaxies in the field are necessary.

In this work we were able to carry out a detailed morphological study of the isolated
CIG$\cap$SDSS sample. Most of E galaxies in this sample show actually fine structures 
(e.g., shells, rings, ripples), dust lanes, ``diffuse haloes'', and inner disks, as well as disky or 
boxy isophotes. The finding of these morphological distortions is based on the judgment of the 
images after different  kernel  sizes were used in the filtering process of E/S0 galaxies as explained in \S  3.1
Notice that we did not applied a quantitative procedure 
for determining distortions with respect to a model light distribution (e.g., Colbert
et al. 2001; van Dokkum 2005). 
The distortions that we call ``diffuse haloes''  (DH) consist of faint diffuse envelopes that
appear symmetric in shape.  It is not clear at this moment 
what is the nature of this component. CIG 533 is an E galaxy showing such a diffuse component. 
Interestingly CIG 533 is the most luminous galaxy ($M_{r} = -22.74$) in our list. Notice however 
that the presence of a nearby (projected on the sky) dwarf galaxy, makes CIG 533 to appear in 
the list of possible interacting galaxies by Sul06.  It should be mentioned that
extended luminous halos were found in numerical simulations of galaxy
formation in the cosmological context; these faint structures consist of stars shed by 
merging subunits during the many accretion events that characterize the hierarchical 
assembly of galaxies (Abadi, Navarro \& Steinmetz 2006).

An E galaxy was considered to be boxy or disky if the $A_{4}$ parameter amplitude reached at 
least a $\pm$ 2.5\% level within the observed radius. This is a first order estimate and it 
should be noticed that no normalization to a characteristic radius (e.g., the effective radius, 
$r_{e}$) is reported.  Examples of boxy and disky isophotes can be seen in CIG 264 and CIG 768, 
respectively. In CIG 245  a shell can be appreciated and in CIG 582, dust lanes. In this 
last case the dust lane appears to affect the fourth--order cosine profile. A localized knot 
near the central region can also be appreciated. These features together suggest 
evidence in favor of a merger event.

\placetable{tbl-6}
 
According to our morphological revaluation, in the CIG$\cap$SDSS sample only 3.5\% of the 
galaxies are of E type, showing that the formation of E galaxies in low--density 
environments is a rare process. The average $r$-band luminosity of the E galaxies $<-21.24 \pm 1.43 >$ 
confirm the scarcity of luminous E galaxies in isolated environments  (e.g., Park et al. 2007).
Table 6 lists the E galaxies in this work and their 
distorted morphology properties.  Column (1) gives the CIG number, Column (2) indicates the 
boxy/disky nature, Column (3) indicates the presence of fine structure in the form  shells/ripples or rings, Column (4) 
indicates the presence of dust lanes, Column (5) the presence 
of inner disks, Column (6) the presence blue/red knots, Column (7) the presence of a diffuse halo $DH$, Column (8) gives 
the ($g-i$) color and Column (9) give the $r$-band absolute magnitude ($H_{0} = 70\, km s^{-1} Mpc^{-1}$).    
Out of the 18 E CIG$\cap$SDSS galaxies, at least 15 (83\%) are characterized by some kind of 
morphological distortion: 50\% show fine stuctures, 17\% present dust lane 
structures, 39\% show a kind of diffuse halo, and 39\% present evidence of an inner
disk; in most of the cases a given galaxy presents combinations of several  of these distortions.
On the other hand, we do not find for most of these E galaxies evidence of blue clumpy 
tidal tails, which are typical in S--S interactions, as well as of other kind of ``smooth''
faint tidal features. It is possible that in deeper images, the latter features 
could appear (see van Dokkum 2005), so that the 83\% of E galaxias with 
morphological distortions found here is actually a lower limit. Notice in Table 6 
the incidence of diffuse haloes ($DH$) in some of the E galaxies. We are not clear at this moment 
about the nature of this component or about it relationship to a merger event.

On the other hand,
we have found that among the 18 E galaxies, 8 are boxy and 7 disky. Interesting enough,
among the 4 galaxies with blue colors (tentatively $g-i<1.0$, see Fig. 7), 3 are disky, 
and the other one is boxy, but it shows evidence of an inner disk.

According to simulation results, E--E (dry) mergers do not develop prominent clumpy tidal tails
and instead are characterized by off--center outer envelopes and the ejections of 
stars (e.g., Combes et al. 1995), which can be seen as fans, shells, rings, envelopes, and tidal 
features. This and other works suggest that the perturbation features in the E--E mergers are of
short duration (less than 0.5 Gyr), so that their detection implies that the merger
happened relatively recently.  We look for the presence of $m=1$ asymmetries from a 
Fourier analysis and found no significant amplitudes for the 18 E galaxies  in this study.

The numerical simulations also show  that the outcome 
of merging early--type galaxies is an anisotropic, slowly rotating, boxy spheroid 
(Naab et al. 2006), while low-mass, highly rotating, disky spheroids are produced typically 
in unequal--mass spiral mergers (Naab et al. 1999; Naab \& Burkert 2003). 
In general, in order that an E system forms from S--S mergers, the merger should be major 
(approximate mass ratios in the range 1:1--3:1; e.g., Bournaud, Jog \& Combes
2005, and more references therein). The latter authors have found that in the case of 
intermediate mergers, with mass ratios in the range 4:1--10:1, hybrid systems that could be
considered as candidates for S0 galaxies form. In the case of minor mergers, the simulations
show that the disturbed disks survive (e.g., Velazquez \& White  1999). 

In conclusion, we can speculate that most of the isolated E galaxies analyzed here suffered 
recent major mergers, mostly dry. However, for a few galaxies we find also pieces of 
evidence of intense knots, central blue colors, in which case a 
major merger could be expected to have involved at least one disk galaxy. Findings of late SF 
in isolated E galaxies were also reported by other authors (Treu et al. 2005; Reda et al. 2007). 
For example, Reda et al. (2007) estimated the 
age of the stellar populations of a sample of local isolated E's revealing that several  
galaxies have central young stars, which also require a recent gaseous accretion or merger.
They conclude that a formation scenario of a single dissipative collapse can not explain 
the spatial distribution of the stellar population and kinematic properties of their studied 
E isolated galaxies. The main question is now in the statistics: what fraction of isolated
E's show evidence of recent ``dynamical'' evolution by dry major mergers and what fraction
corresponds to E's with recent SF, product of mergers of galaxies with gas
and/or late gas accretion. For the small sample studied here, the former seems to be
the case. The above question is relevant for the models of galaxy formation
and evolution.  Studies with larger samples are needed.

While there is significant evidence that E galaxies could be formed by merging scenarios, 
the origins of S0 galaxies are more obscure. Since most S0 studies have concentrated on 
cluster S0 galaxies, the proposed origins of these systems mostly involve processes that 
remove gas from disk galaxies and thus truncate SF (Mihos et al. 1995; Quilis et 
al. 2000). However, some of these processes may not be relevant to the histories of very 
isolated S0 galaxies. The relative numbers of E and S0 galaxies in the CfA survey is 1:4 
in favor of S0 galaxies (Marzke et al. 1994). This is in contrast to their relative abundance 
in the CIG Catalog, 1:1.15  (Sul06) and 1:1.85 (this work), suggesting this the
importance of the environment on the formation of S0 galaxies. A possible way to form S0's
in an isolated environment has been mentioned already above: galaxy mergers of intermediate
mass ratios, 4:1--10:1 (Bournaud et al. 2005).

\begin{figure}
\plotone{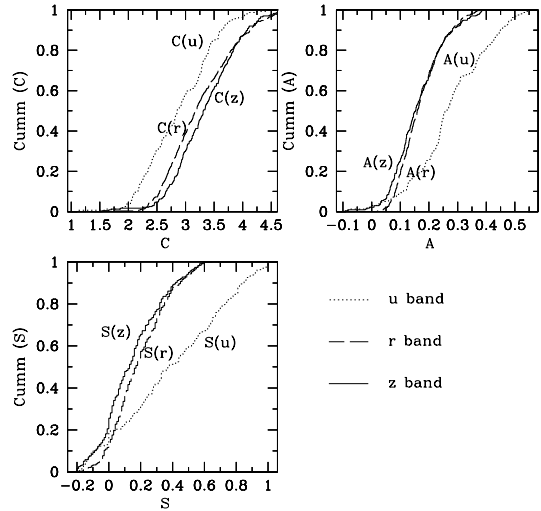}
\caption{Cumulative distribution function of the $CAS$ parameters for CIG galaxies at $u$, $r$ and $z$ bands. }
\label{mosaico1}
\end{figure} 

\begin{figure}
\plotone{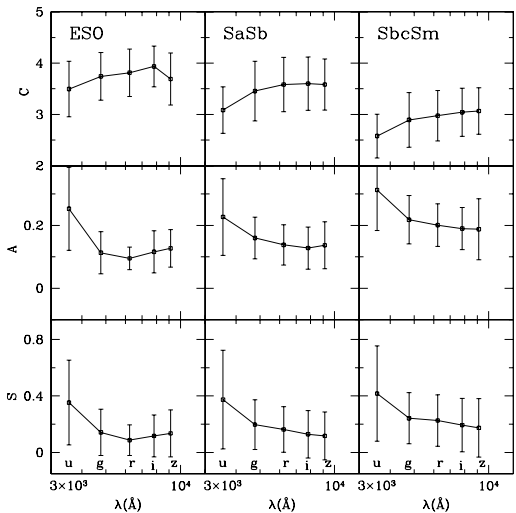}
\caption{Average and standard deviation values of the $CAS$ parameters vs SDSS band for CIG galaxies 
sorted in to E/S0 types (left panel), SaSb types (middle panel) and SbcSm types (right panel).}
\label{mosaico1}
\end{figure}

\subsection{Physical Morphology}
 
We start by exploring how the $CAS$ parameters for CIG galaxies do change with wavelength.  Figure 10 shows the cumulative distribution 
function of the $CAS$ parameters at $u$, $r$ and $z$ bands.  Significant changes in the $CAS$ parameters with wavelength are appreciated. 
The concentration $C$ becomes higher from bluer to redder bands while in the case of both the asymmetry $A$ and clumpiness $S$ parameters, 
their values strongly decrease from bluer to redder bands.

In Figure 11 we plot the average and standard deviation values of the $CAS$ parameters vs wavelength (color band) for the CIG galaxies sorted 
into E/S0, early-- and late--type spirals (E/S0 --left panel-- SaSb --middle panel-- and SbcSm --right panel, respectively). The $CAS$ parameters 
of later types show on average more dependence with wavelength than the early types, as well as more scatter at a given band.  Among the 
$CAS$ parameters, the clumpiness is the most sensitive to wavelength, although the corresponding error bars indicate a significant dispersion 
in this parameter.
 
\begin{figure*}
\plotone{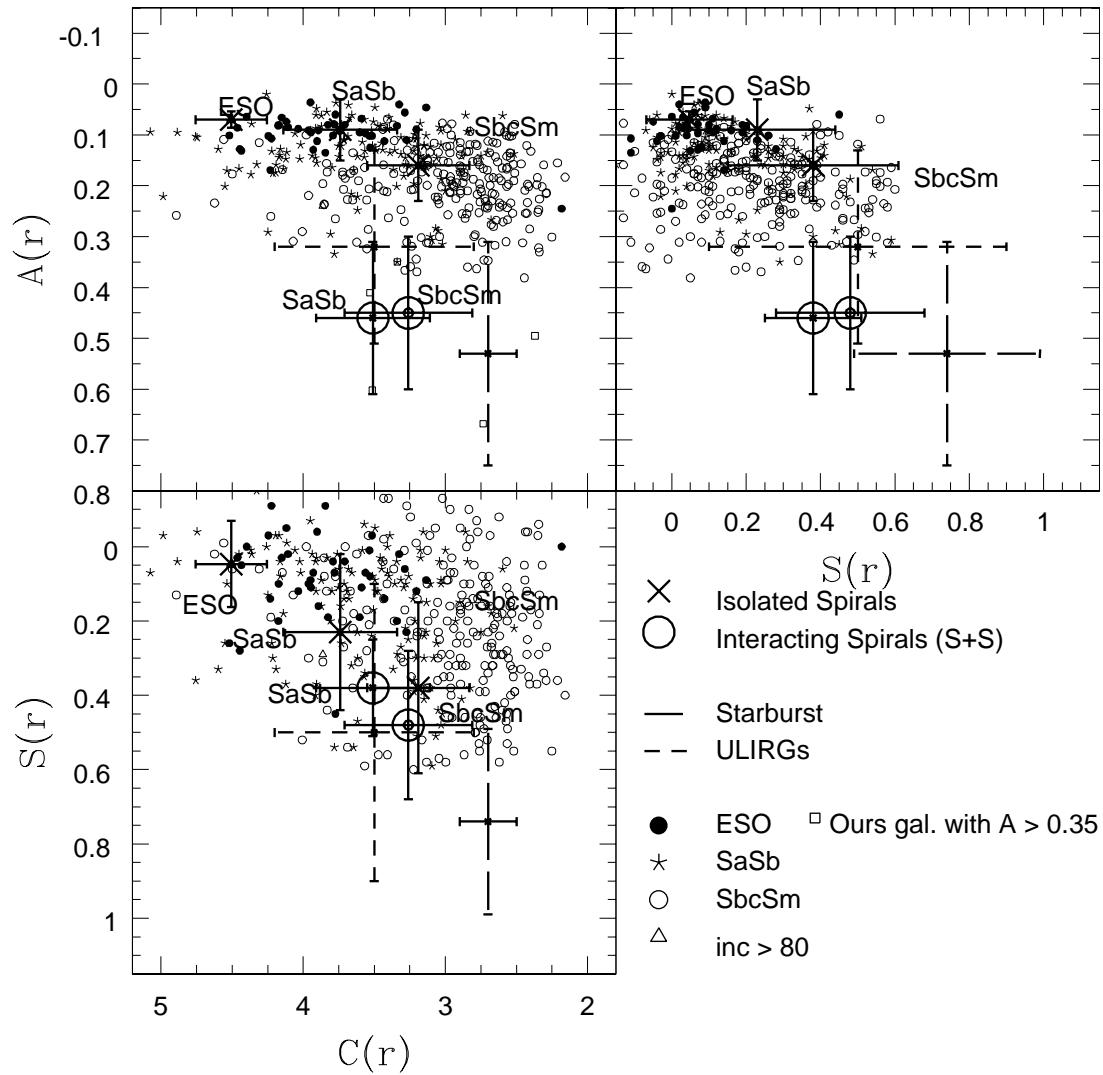}
\caption{The loci of the CIG$\cap$SDSS sample in the $CAS$ diagrams. For comparison, the averages and standard deviations of SaSb and SbcSm 
subsamples of CIG galaxies in the Johnson-Cousins $R$-band system (Hern\'andez-Toledo et al. 2007; 2008) and of E/S0 isolated galaxies 
from the HyperLeda database (Hern\'andez-Toledo et al. 2006)  are shown (crosses and continuous error bars). The $R-$band averages and standard 
deviations of galaxies in interacting S+S pairs (Hern\'andez--Toledo et al. 2005),  and Starburst and Ultra Luminous 
Infrared (ULIR) galaxies (C2003) are also plotted. }
\label{mosaico1}
\end{figure*} 

Figure 12 shows the loci of the CIG$\cap$SDSS galaxies sorted into E/S0, SaSb and SbcSm galaxies in the projected planes of the $r-$band $CAS$ 
space.  For comparison,  the averages and standard deviations of SaSb and SbcSm subsamples of CIG galaxies in the Johnson-Cousins $R$-band  
system (Hern\'andez-Toledo et al. 2007; 2008) and of E/S0 isolated galaxies from the HyperLeda database (Hern\'andez-Toledo et al. 2006)  are shown 
(crosses and continuous error bars). Also the $R-$band averages and standard deviations of galaxies in interacting S+S pairs (Hern\'andez--Toledo et al. 
2005),  and Starburst and Ultra Luminous Infrared (ULIR) galaxies (C2003) are also plotted. 


The question of whether the disk of a galaxy is intrinsically asymmetric or not is of great interest. Some studies have shown that important 
deviations from axisymmetry exist in the optical and other wavelengths (Rix \& Zaritsky 1995; Richter \& Sancisi 1994; C2003). However, systematic 
attempts to quantify asymmetry and other measures like the $CAS$ parameters in several wavelengths for well--selected local samples of isolated 
galaxies are rare and missing in the literature.

The asymmetry parameter $A$ has been shown to be sensitive mainly to galaxy interactions/mergers, but also is influenced by SF clumps, dust 
lanes, and projection effects. The quantitative measure of $A$ in the present sample of CIG$\cap$SDSS galaxies spans roughly the range $0.0 \leq A(r) \leq 0.4$, 
the average and standard deviation being $<A(r)> = 0.17\pm 0.07$. The later types are slightly more asymmetric on average than the earlier ones. 
The mean asymmetries reported here are definitively lower than the typical ones of interacting disk galaxies (Hern\'andez--Toledo et al. 2005).  

Next, we  illustrate how the $CAS$ parameters of the CIG galaxies correlate with other properties 
and whether these correlations are sensitive to 
the passband or not. Figure 13 shows the $griz$ band $CAS$ parameters vs morphological type $T$.
Code numbers  for each morphological type are now associated according to the HyperLeda convention. 
The trend seen of the $CAS$ parameters is a consequence of the natural flocculency and SF in 
galaxies as they get later in type. SF is more active for later types (as is evidenced also by 
the trend of higher $A$ and $S$ values as the types are later and the color bluer). 
 
The correlations of the $CAS$ parameters with $T$ are very scattered, the tightest
being with the concentration parameter $C$ in the sense of smaller values of $C$ as the 
type is later. For the $A$ parameter, one sees only a weak trend of $A$ on average being 
higher as the type is latter, and for the $S$ parameter, it is not possible to define even a 
trend.  The trends of $C$ and $A$ with $T$ are more robust in the $r$ and $i$ bands,
suggesting this that the basic structure of galaxies is better revealed in the infrared
bands.

 Figures 14 and 15 show the $CAS$ parameters vs the total $(g-i)$ color for the CIG sample 
sorted into E/SO and Sa-Sm types, respectively. Corrected colors are presented for Sa-Sm types. Nearly edge--on galaxies (inclination $\ge 80^{\rm o}$) 
are plotted with crosses. Given that $A$ and $S$ are particularly sensitive to projection 
effects, it is important to visualize these galaxies since they may be masking any trend.


 \begin{figure}
\plotone{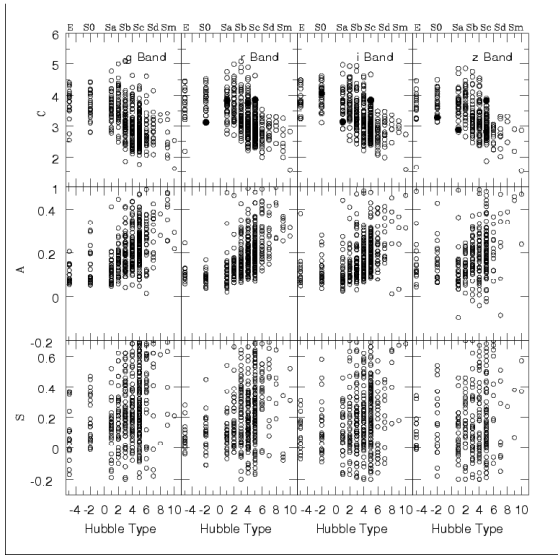}
\caption{The structural $CAS$ parameters vs the Hubble type for the CIG sample sorted into SDSS band. First column: $g$-band, 
second column: $r$-band, third column: $i$-band and fourth column: $z$-band.}
\label{mosaico1}
\end{figure}

Figure 14 shows the natural scatter in the $CAS$ parameters and a segregation in blue and red colors for the isolated 
E/S0 galaxies (Hern\'andez-Toledo et al. 2006).  Figure 15 shows the higher sensitivity of the $CAS$ parameters to intrinsic color variations in 
isolated spiral galaxies  (Hern\'andez-Toledo et al. 2005). 

Although it is expected that the color trends are more robust towards the redder bands  which are less contaminated from (transient)  SF effects 
thus better representing the basic structure of galaxies, we find a large scatter in the $z$ band. This is indicating the intrinsically noisier nature 
of this band. 

\begin{figure}
\plotone{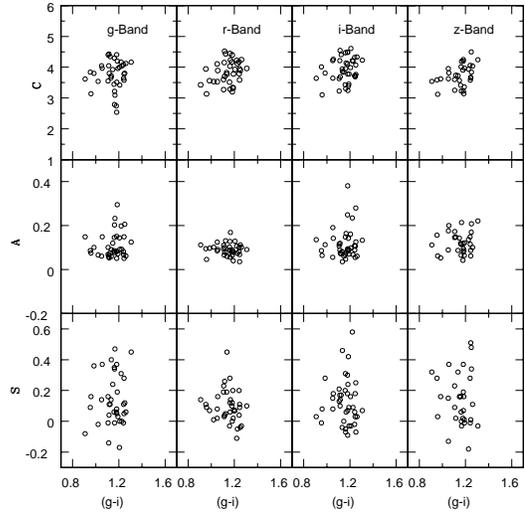}
\caption{The structural $CAS$ parameters vs the $(g-i)$ color for E/S0 CIG galaxies, sorted into SDSS band. First column: $g$-band, 
second column: $r$-band, third column: $i$-band and fourth column: $z$-band.}
\label{mosaico1}
\end{figure} 

\begin{figure}
\plotone{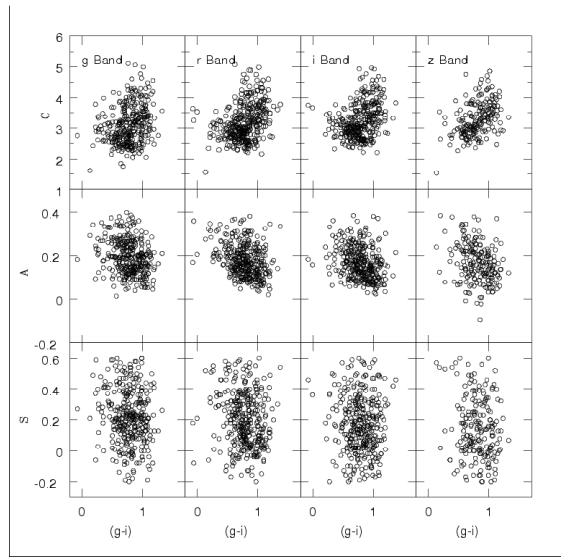}
\caption{The structural $CAS$ parameters vs the $(g-i)$ color for SaSm CIG galaxies. Same as in Figure 14.}
\label{mosaico1}
\end{figure} 

\section{Summary and Conclusions} \label{S5}

The building up of a well--defined sample of local isolated galaxies with uniformly--derived and 
detailed morphological information is of great relevance because it provides a fair database 
for comparison with model predictions as well as with observed samples of galaxies in 
other environments and at higher redshifts. In this paper we reported (i) a detailed morphological 
revaluation of a sample of 549 CIG galaxies that were available in the SDSS(DR6) database,
(ii) looked for morphological distortions in the analyzed isolated galaxies, (iii) calculated
corrected magnitudes and colors, as well as the structural $CAS$ parameters for all the galaxies, 
and (iv) discussed some preliminary conclusions based on the obtained results.
  
We have taken advantage of the improved scale and dynamic range of the SDSS database and 
proposed an image processing scheme that enhances the detection of low/high surface brightness 
morphological and structural components in these CIG galaxies. We have used logarithmic scaled 
images to look for internal/external details; $g$ band filtered--enhanced images to look for 
internal structure in the form of star forming regions, bars, rings and/or structure embedded 
into dusty regions; and used complementary RGB color images from the SDSS database to visualize 
the spatial distribution of the SF and other components like dust (blue colors for recent SF 
and red colors for older populations/dusty components). 

In the case of E/S0/Sa galaxy candidates, we have estimated and used additional information in the 
form of a surface brightness profile and the corresponding geometric profiles (ellipticity 
$\epsilon$, Position Angle $PA$ and $A_{4}/B_{4}$ coefficients of the Fourier series expansions 
of deviations of a pure ellipse) from the $r$ band images in order to provide evidence of 
boxiness/diskiness and other structural details that help to discriminate morphology in 
these galaxies.

The main results and conclusions of the paper are as follow: 

--The median morphological type in the present CIG$\cap$SDSS sample is Sbc ($T=4$); 
35\% of the galaxies are earlier than Sbc, and 65\% are Sbc or later. 
This new revaluation gives somewhat different morphological fractions with respect to
previous reports, mainly in the E, S0, Sa and Sab types. While Sul06 found 5.7\% of E types, 
6.6\% of S0 types and 1.4\% of transition E/S0 types, amounting to 
13.7\% of E/S0 galaxies, our results indicate 3.5\% of E types and 
5\% of S0 types, amounting to a fraction of only 8.5\% of E/S0 galaxies. This 
emphasizes the difficulty of finding E/S0 galaxies in environments typical to the isolated 
galaxies. For early--type S galaxies, while Sul06 found 1.3\% 
of Sa types and 5.1\% of Sab types, we accounted for 6.5\% of Sa types and 7.8\% of Sab types. 
In general, for 74\% of the galaxies in common with Sul06 both classifications
have a difference $|\Delta T | < 2$. 

--For the 539 CIG$\cap$SDSS galaxies we have calculated the absolute magnitudes in the bands 
$gri$ and corrected by both the Galaxy and internal extinction. The average and 
1$\sigma$ dispersions of the $g-i$ color for the E/S0, Sa--Sb, and Sbc--Sm/Irr morphology 
ranges are $1.18 \pm 0.26$, $0.89 \pm 0.21$, and $0.66 \pm 0.55$, respectively. The color--magnitude
diagram for the E/S0's is quite flat and narrow, while for the disk galaxies, a 
trend of redder galaxies as they get brighter, but with a large scatter is appreciated. 

--We reported a tentative $r-$band fraction of bars in disk galaxies of 65.8\% 
(defined only by an eye inspection and excluding objects with $i> 60^{\circ}$); $35\%$ are 
clearly barred (SB), and $30.8\%$ show evidence of a weak bar (SAB). The reported fraction
of bars in the present CIG isolated disk galaxies is larger than others reported in the literature 
for samples related to field and cluster/group environments. The fraction of barred galaxies
does not change with the morphological type, but it correlates significantly with the light
concentration parameter $C_{80/20}$. 
We have also found 33.3\% of the disk galaxies showing ring structures.

--Our image processing scheme of the SDSS data allowed us to find a richness of distinct 
substructure in the isophotal shapes of the E galaxies as well as a series of morphological
distortions. Out of the 18 E galaxies in the sample, 8 (44\%) show boxy isophotes and 7 (38\%) 
show disky isophotes. Among the 4 blue E's ($g-i < 1$), 3 are disky and 1 boxy, but the latter
shows evidence of an inner disk. The fraction of E's with one or more kinds of morphological 
distortion amounts to 78\%: fine stuctures (50\%), dust lane structures (17\%), 
diffuse halo (39\%), and inner disk (39\%). Our results suggest that most of the isolated E's 
suffered recent dry major mergers (E--E). Only a few of them show pieces of evidence of mergers 
of galaxies with gas, and/or of late gas accretion and intense SF.

--In order to complement our morphological study, we have estimated the $ugriz$ band 
$CAS$ parameters of the 549 CIG$\cap$SDSS galaxies as part of a model--independent 
morphological system. The $CAS$ averages and standard deviations in the $r-$band for the 
E/S0 galaxies are $<C> = 3.81 \pm 0.46$, $<A> = 0.09 \pm 0.03$, $<S> = 0.08 \pm 0.10$, 
for the SaSb galaxies are $<C> = 3.58 \pm 0.52$, $<A> = 0.14 \pm 0.06$, $<S> = 0.16 \pm 0.16$,
and for the SbcSm galaxies are $<C> = 2.97 \pm 0.49$, $<A> = 0.20 \pm 0.07$, $<S> = 0.25 \pm 0.18$. 
These values are in rough agreement with those reported previously for smaller 
samples of isolated galaxies in the $R$-band Johnson-Cousins system 
(Hern\'andez-Toledo et al. 2006;2008).  The mean asymmetries reported here 
are definitively lower than the typical ones for interacting disk galaxies 
(Hern\'andez--Toledo et al. 2005).  
While the $C$ parameter systematically increases from bluer to redder bands, both $A$ and $S$ 
significantly decrease. The $CAS$ parameters present more robust trends with the morphological 
type $T$ and the total $(u-r)$ color in the redder bands, suggesting that the basic structure 
of galaxies is revealed better in the redder bands. In any case, the mentioned trends are very 
noisy, the less scattered being with $C$.

\begin{acknowledgements}

H.M.H.T. thanks the staff of the Observatorio Astron\'omico Nacional in
San Pedro M\'artir, B. C. for the help in the observations of subsamples of the CIG in the 
Johnson-Cousins photometric system. The authors thank Changbom Park for fruitful discussions.
This work was funded by CONACYT grant 42810 to H.M.H.T, J.A.V.M. and S.O.E. We thank referee Michael 
S. Vogeley for his comments/suggestions that improved the content of this manuscript. 

This research has made use of the NASA/IPAC Extragalactic Database (NED) which is 
operated by the Jet Propulsion Laboratory, California Institute of Technology, 
under contract with the National Aeronautics and Space Administration. 

We acknowledge the usage of the HyperLeda database (http://leda.univ-lyon1.fr).
 
Funding for the SDSS and SDSS-II has been provided by the Alfred P. Sloan Foundation, the Participating Institutions, the National Science Foundation, the U.S. Department of Energy, the National Aeronautics and Space Administration, the Japanese Monbukagakusho, the Max Planck Society, and the Higher Education Funding Council for England. The SDSS Web Site is http://www.sdss.org/.

 The SDSS is managed by the Astrophysical Research Consortium for the Participating Institutions. The Participating Institutions are the American Museum of Natural History, Astrophysical Institute Potsdam, University of Basel, University of Cambridge, Case Western Reserve University, University of Chicago, Drexel University, Fermilab, the Institute for Advanced Study, the Japan Participation Group, Johns Hopkins University, the Joint Institute for Nuclear Astrophysics, the Kavli Institute for Particle Astrophysics and Cosmology, the Korean Scientist Group, the Chinese Academy of Sciences (LAMOST), Los Alamos National Laboratory, the Max-Planck-Institute for Astronomy (MPIA), the Max-Planck-Institute for Astrophysics (MPA), New Mexico State University, Ohio State University, University of Pittsburgh, University of Portsmouth, Princeton University, the United States Naval Observatory, and the University of Washington.

\end{acknowledgements}


\end{document}